\documentclass[conference]{IEEEtran}
\usepackage{listings}
\usepackage{float}
\usepackage[ruled,vlined]{algorithm2e}
\usepackage{url}
\usepackage{hyperref}
\usepackage{tikz}
\usepackage{amsmath}
\usepackage{booktabs}
\usepackage{amssymb}
\usepackage{multirow}
\usepackage{filecontents}
 \usepackage{subcaption}
\usepackage{caption}
\usepackage{stfloats}    % for table* at top

\AtBeginDocument{%
  }

\begin{document}

%%
%% The "title" command has an optional parameter,
%% allowing the author to define a "short title" to be used in page headers.
\title{Redefining Website Fingerprinting Attacks\\with Multi-Agent LLMs}

%%
%% The "author" command and its associated commands are used to define
%% the authors and their affiliations.
%% Of note is the shared affiliation of the first two authors, and the
%% "authornote" and "authornotemark" commands
%% used to denote shared contribution to the research.

% \author{First Last}
% \authornote{Both authors contributed equally to this research.}
% \email{trovato@corporation.com}
% \orcid{1234-5678-9012}
% \author{G.K.M. Tobin}
% \authornotemark[1]
% \email{webmaster@marysville-ohio.com}
% \affiliation{%
%   \institution{Institute for Clarity in Documentation}
%   \city{Dublin}
%   \state{Ohio}
%   \country{USA}
% }

% \author{Chuxu Song}
% \affiliation{%
%   \institution{Rutgers University}
%   \city{New jersey}
%   \country{US}}
% \email{cs1346@scarletmail.rutgers.edu}

% \author{Dheekshith Dev Manohar Mekala}
% \affiliation{%
%   \institution{Rutgers University}
%   \city{New jersey}
%   \country{US}}
% \email{}

% \author{Hao Wang}
% \affiliation{%
%   \institution{Rutgers University}
%   \city{New jersey}
%   \country{US}}
% \email{hw488@cs.rutgers.edu}

% \author{Richard Martin}
% \affiliation{%
%   \institution{Rutgers University}
%   \city{New jersey}
%   \country{US}}
% \email{rmartin@scarletmail.rutgers.edu}
\author{\IEEEauthorblockN{Chuxu Song}
\IEEEauthorblockA{Rutgers University\\
cs1346@scarletmail.rutgers.edu}
\and
\IEEEauthorblockN{Dheekshith Dev Manohar Mekala}
\IEEEauthorblockA{Rutgers University\\
dm1653@scarletmail.rutgers.edu}
\and
\IEEEauthorblockN{Hao Wang}
\IEEEauthorblockA{Rutgers University\\
hw488@cs.rutgers.edu}
\and
\IEEEauthorblockN{Richard Martin}
\IEEEauthorblockA{Rutgers University\\
rmartin@scarletmail.rutgers.edu}}

\newcommand{\hao}[1]{\textcolor{red}{[Hao: #1]}}

%%
%% This command processes the author and affiliation and title
%% information and builds the first part of the formatted document.
\maketitle
\begin{abstract}
Website Fingerprinting (WFP) uses deep learning models to classify encrypted network traffic to infer visited websites. While historically effective, prior methods fail to generalize to modern web environments. Single-page applications (SPAs) eliminate the paradigm of websites as sets of discrete pages, undermining page-based classification, and traffic from scripted browsers lacks the behavioral richness seen in real user sessions. Our study reveals that users exhibit highly diverse behaviors even on the same website, producing traffic patterns that vary significantly across individuals. This behavioral entropy makes WFP a harder problem than previously assumed and highlights the need for larger, more diverse, and representative datasets to achieve robust performance. To address this, we propose a new paradigm: we drop session-boundaries in favor of contiguous traffic segments and develop a scalable data generation pipeline using large language models (LLM) agents. These multi-agent systems coordinate decision-making and browser interaction to simulate realistic, persona-driven browsing behavior at 3--5× lower cost than human collection. We evaluate nine state-of-the-art WFP models on traffic from 20 modern websites browsed by 30 real users, and compare training performance across human, scripted, and LLM-generated datasets. All models achieve under 10\% accuracy when trained on scripted traffic and tested on human data. In contrast, LLM-generated traffic boosts accuracy into the 80\% range, demonstrating strong generalization to real-world traces. Our findings indicate that for modern WFP, model performance is increasingly bottlenecked by data quality, and that scalable, semantically grounded synthetic traffic is essential for capturing the complexity of real user behavior.
\end{abstract}

%%The rise of privacy-preserving technologies such as Tor and Virtual Private Networks (VPNs) has significantly improved users' ability to browse the internet anonymously. Tor, for instance, routes traffic through a series of volunteer-operated relays and applies multiple layers of encryption to obfuscate both the content and the origin of network packets. Similarly, VPNs create encrypted tunnels between users and remote servers, hiding destination IPs and shielding traffic from local surveillance. Despite their effectiveness, these systems remain vulnerable to a powerful class of traffic analysis attacks known as Website Fingerprinting (WFP).

Website Fingerprinting attacks~\cite{old_wfnet,tmwf,tf,critical_evaluation,tor_wang,277132,10.1145/3658644.3690211,298090,explainwf-popets2023,10179464} exploit side-channel information from encrypted traffic to infer which website a user is visiting. Even though content and destination IPs are hidden, traffic patterns such as packet sizes, directions (incoming vs. outgoing), burst lengths, and inter-arrival times can serve as discriminative features for identifying web activity. Early WFP techniques used traditional machine learning classifiers on hand-engineered features~\cite{panchenko2011website, cai2012touching}, but recent work has demonstrated that deep learning models—such as CNNs and LSTMs—can achieve significantly higher accuracy under controlled conditions~\cite{df, varcnn, hayes2016kfingerprinting}. These models typically operate on network traces corresponding to full webpage loads and achieve classification accuracy higher than 90\% in closed-world settings.

\section{Introduction}
\begin{figure}[t]
  \centering
  \includegraphics[width=0.9\columnwidth]{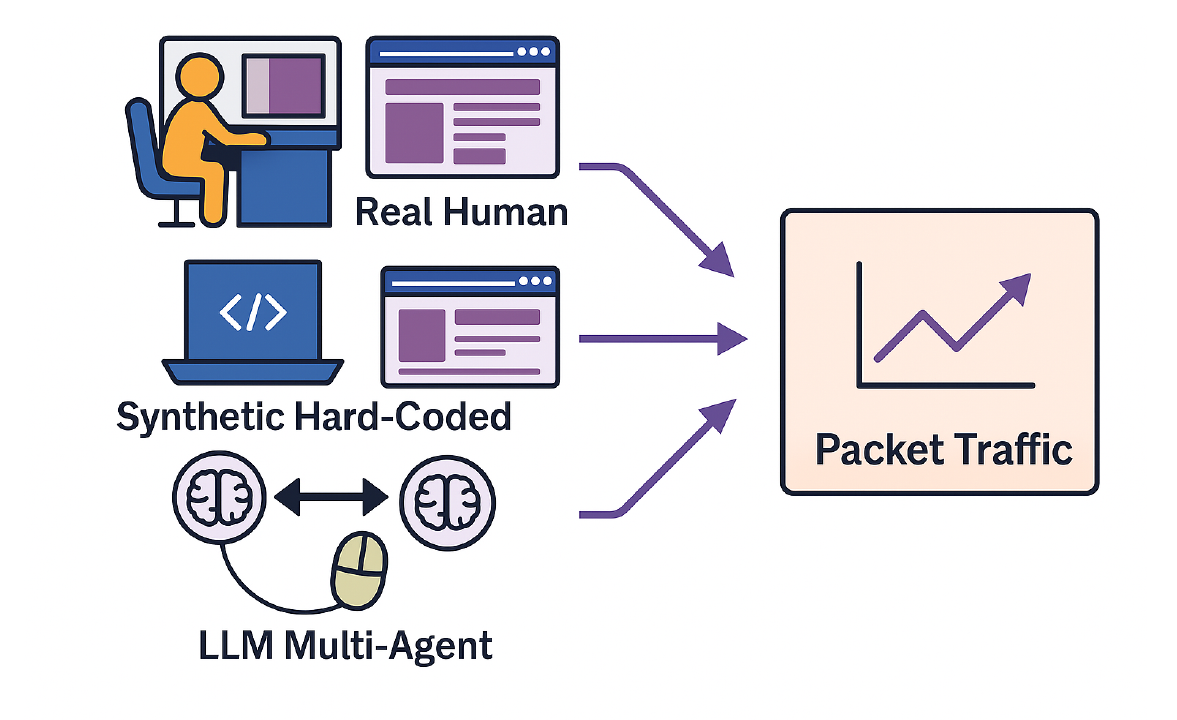}
  \caption{We collect WFP traffic dataset generated by real human users, traditional scripted robots, and our LLM-based multi-agent system.}
  \label{fig:wfp-llm}
\end{figure}
However, most of these techniques implicitly assume a static, page-oriented web browsing model. In this model, users navigate between clearly delineated web pages, and each page load corresponds to a discrete unit of classification.
Although researchers have introduced enhancements such as webpage segmentation~\cite{juarez2016towards}, early-stage prediction~\cite{raptor2017early,cheng2025holmes,deng2024robustreliableearlystagewebsite}, and online learning~\cite{oh2021online}, these methods fail to consider
that many websites have evolved well past the model of a collection of static pages. Today’s websites are dynamic, interactive, and highly personalized. Single Page Applications (SPAs) and streaming platforms dominate user traffic, where content is updated dynamically without changing URLs. On platforms like YouTube, Reddit, or Amazon, users may stay on a single webpage for extended periods, interacting with embedded media players, infinite-scroll feeds, or personalized content panels.
These interactions create rich and varied traffic that breaks the clean boundaries required for traditional WFP analysis.

\begin{figure*}[t]
    \centering
    \includegraphics[width=0.95\linewidth, height=0.23\textheight]{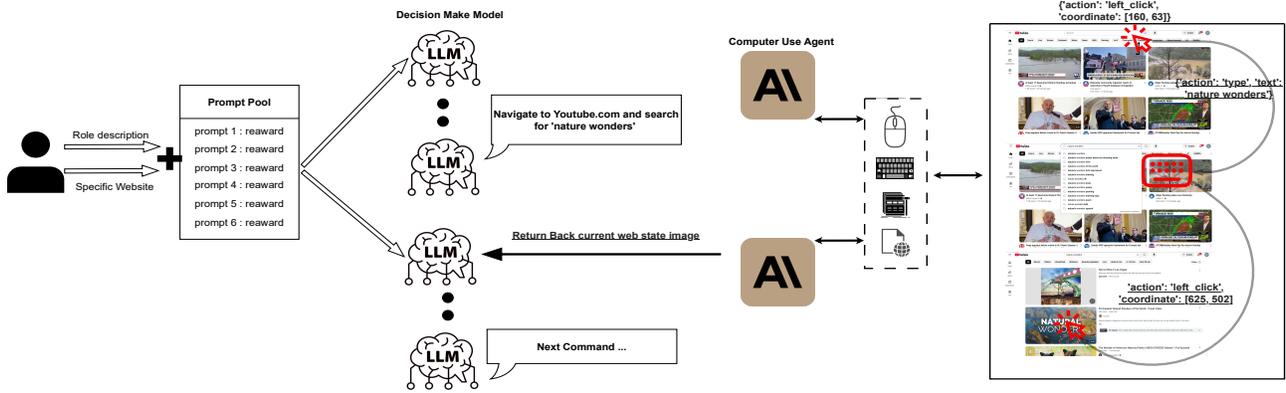}
    \caption{Overview of our proposed LLM-driven multi-agent framework for generating realistic website fingerprinting (WFP) training data. Prompt-conditioned language models simulate goal-oriented user commands, which are executed in a browser environment. The system forms a closed loop by feeding back updated visual states, enabling scalable, behavior-rich, and semantically grounded interaction traces.}
    \label{fig:pipeline}
\end{figure*}

We revisit the WFP threat model in the context of modern, long-lived, and interaction-rich browsing sessions. We begin by addressing a foundational need in WFP research: realistic, high-quality traffic data that accurately reflects human behavior. We hired 30 participants to browse 20 modern websites—including several SPAs and content-dense platforms—on controlled desktop environments. Participants engaged with the sites freely for over 100 hours in total, generating a real-world dataset of encrypted traffic traces. To simulate existing synthetic approaches, we collected network traffic using a scripted browser, which followed same-domain links using a breadth-first crawl without performing any meaningful interaction.

Our analysis reveals a significant representativeness gap between traffic generated by synthetic scripted browsers and human traffic that
impacts WFP classification accuracy across many models. For example, over 60\% of human sessions involved spending more than 10 consecutive minutes on a single webpage, with extensive in-page activity such as scrolling, hovering, typing, or clicking dynamically loaded components. These behaviors produce long, variable-length traces with complex temporal dynamics that defy traditional WFP segmentation methods. 

To simulate realistic attack conditions, we assume the attacker has no access to precise session start points or ground-truth webpage splits. We evaluated eight state-of-the-art WFP models across three training/testing regimes: (1) both training and testing scripted traffic, (2) training on scripted traffic
and testing on human traffic, and (3) training on human traffic and testing on leave-one-out human traffic where we test on a human's traffic
not in the training set. Results show that while human-to-human training achieves up to 75\% accuracy, scripted-to-human adaptation yields
less than 15\% accuracy, highlighting the poor generalization of current synthetic traffic generation methods when directly programming
a browser with scripts. 

Despite the value of human-generated traffic, collecting and scaling such a training corpus is prohibitively expensive and raises serious privacy concerns. Recruiting diverse participants, compensating them fairly, and enforcing consistent browsing protocols requires significant manual effort. Moreover, due to ethical limitations, it's difficult to capture sensitive behavioral patterns or long-term personalized activity.

To improved WFP classification accuracy in the modern Internet, we introduce two novel approaches. First, we abandon the view of
the web as a set of discrete pages. Rather than try to classify web pages, we classify entire web sites; that is, the output
of the models is an entire web site domain rather than any single page. Second, rather than use complex and error prone attempts
to find page boundaries, we only use 5000 packets of contiguous traffic as input to a deep learning model. Continuous traffic observations 
correspond well to a simple attack model where an attacker has gained observational access in a network component, such as an end point 
operating system or network middlebox, and can identify web sites by their destination IP addresses.

To address the question of how to get representative traffic, we propose a novel traffic generation pipeline based on a multi-agent architecture~\cite{talebirad2023multiagentcollaborationharnessingpower, ijcai2024p890, Li_Wang_Zeng_Wu_Yang_2024, wu2023autogenenablingnextgenllm, hu2025positionresponsiblellmempoweredmultiagent} powered by Large Language Models (LLMs). An overview of our architecture is illustrated in Figure~\ref{fig:pipeline}, which showcases the interaction between two key components: a high-level \textit{Decision-Making Agent} and a low-level \textit{Computer-Using Agent} (CUA). The \textit{Decision-Making Agent} is responsible for generating semantically meaningful, context-aware navigation instructions based on the current webpage state, while the \textit{Computer-Using Agent (CUA)} performs fine-grained browser operations—including clicking, scrolling, typing, and DOM navigation—based solely on high-level natural language commands from the Decision-Making Agent.

The \textbf{Decision-Making Agent} leverages multimodal LLMs such as Claude or GPT to produce concise, screen-grounded commands conditioned on user personas. Given a screenshot and site context (provided by the CUA), it interprets page content and generates navigational intents like “scroll for more articles” or “click the first product,” reflecting realistic and goal-oriented user behavior.

The \textbf{Computer-Using Agent (CUA)} executes these commands inside a real browser using Claude’s Computer Use API, which unifies perception and action within a single LLM interface. Internally, the CUA parses the visual page structure, selects UI elements based on semantics, and emits interaction plans in JSON format compatible with browser instrumentation. This enables high-fidelity, context-aware GUI actions without relying on hand-coded scripts.

By decoupling decision-making from execution, our multi-agent pipeline enables scalable, cross-site browsing simulation with fine-grained behavioral control—without requiring model fine-tuning or rule-based automation. To promote realism and diversity, each Decision Agent is conditioned on a distinct user persona—such as a first-time visitor, a price-sensitive shopper, or a news reader with targeted interests. These personas influence not only the types of actions generated, but also their frequency, depth, and semantic coherence, yielding rich, temporally grounded interaction patterns. As a result, the synthetic traffic produced by our system better reflects the variability of real-world browsing compared to traditional scripted approaches.

We integrate this synthetic traffic into WFP training workflows and observe substantial improvements in generalization. When models are trained jointly on human and LLM-generated traces, cross-user accuracy improves by over 50\%. Even when trained solely on LLM traffic, models achieve up to 3$\times$ higher accuracy on human traces than those trained on classical synthetic datasets. These results demonstrate that high-fidelity LLM-driven simulation—guided by semantic intent and behavioral diversity—can significantly narrow the realism gap in website fingerprinting tasks.

\subsection*{Contributions}

In summary, this paper makes the following contributions:
% \hao{Can we list our collected LLM-based data as the fourth contribution, especially when the size of the dataset is large enough. We may shrink the Paper Organization paragraph is more space is needed.}
\begin{itemize}
    \item We collect and characterize a large-scale dataset of encrypted web traffic from 30 human participants browsing 20 dynamic, modern websites over 100 hours of interaction. The dataset totals over 90GB of human-generated traffic, with approximately 4.5GB per site.
    %% and was gathered at a cost of \$2000 in participant compensation and supervision.
    \item We demonstrate that state-of-the-art WFP models severely overestimate attack effectiveness when trained only on scripted browsers, especially in the context of modern single-page or streaming websites.
    \item We introduce a scalable, LLM-based multi-agent framework for generating realistic synthetic network traffic that significantly improves WFP model generalization for diverse human behavior. 
    \item We construct a complementary dataset using LLM agents covering 10 websites and including 10,000 training samples, offering a scalable and cost-effective alternative to traditional traffic collection.
    %% powered by \$600 of Claude API credits.
\end{itemize}

% version 1
% \paragraph{Paper Organization.}
% Section~\ref{sec:background} introduces key observations from empirical experiments and motivates the need for revisiting Website Fingerprinting (WFP) under modern web conditions. Section~\ref{sec:threatmodel} formalizes a stricter threat model reflecting realistic deployment assumptions, including the lack of clean session boundaries and user-level segmentation. Section~\ref{sec:methodology} outlines our data collection methodology, covering real human browsing traces, traditional synthetic robot traffic, and a scalable multi-agent system using large language models (LLMs). We further detail the prompt optimization framework and modular agent architecture that enable semantically grounded interactions. Section~\ref{sec:experiments} presents comprehensive empirical results across four dimensions: (1) baseline performance under controlled and realistic browsing data; (2) cross-user and cross-domain generalization; (3) the effectiveness of LLM-generated traffic compared to synthetic and human baselines; and (4) the impact of dataset scale on transfer performance. Section~\ref{sec:relatedwork} reviews prior WFP attacks, defenses, and relevant research on agent-based simulation and prompt engineering. We conclude in Section~\ref{sec:conclusion} with a summary of findings and opportunities for future work in scalable, behavior-driven WFP research.

\paragraph{Paper Organization.} 
Section~\ref{sec:background} motivates the need to revisit Website Fingerprinting (WFP) under modern web conditions. Section~\ref{sec:threatmodel} presents our updated threat model. Section~\ref{sec:methodology} details our data collection pipeline and LLM-based simulation framework. Section~\ref{sec:experiments} evaluates model generalization across various training sources and data scales. Section~\ref{sec:relatedwork} reviews prior work. Section~\ref{sec:conclusion} concludes with insights and future directions.

\section{Background and Motivation}
\label{sec:background}

Website Fingerprinting (WFP) is a passive traffic analysis technique that attempts to infer the websites a user is visiting by analyzing encrypted network metadata such as packet sizes, directions, and inter-packet delays. Prior research has demonstrated strong performance using deep learning-based classifiers under closed-world assumptions, achieving over 90\% accuracy on datasets with clear session boundaries and static website structures~\cite{panchenko2011website, hayes2016kfingerprinting, df, raptor2017early}.

However, these results rely on several simplifying assumptions: browsing traces correspond to full page loads; session start and end points are known; and user behavior is deterministic or scripted. These assumptions are increasingly incompatible with modern web usage patterns.

Today’s websites—such as YouTube, Amazon, TikTok, and Reddit—adopt single-page application (SPA) architectures. Users often remain on a single page while triggering asynchronous content updates through scrolling, searching, or media playback. These behaviors generate long, unsegmented, and high-entropy traffic streams, which traditional WFP models were never designed to handle. Compounding this, user interactions vary widely in intent and style, introducing significant intra-site and intra-user diversity.

In our user study (Section~\ref{sec:methodology}), we observed that participants frequently spent over 10 minutes continuously browsing within the same website context, without changing the page or triggering explicit reload. In such sessions, the assumption of “webpage-level splitting” becomes invalid: there are no clear structural boundaries between tasks, actions, or content states. Furthermore, many modern websites adopt proactive content preloading, where elements such as video feeds, product panels, and comment sections are fetched in the background in anticipation of user engagement. This design strategy obfuscates the semantic start point of user behavior, making it exceedingly difficult for an attacker to detect the true beginning of a session—even if segmentation were possible.

To better understand how these behavioral and architectural shifts affect WFP, we benchmarked nine state-of-the-art classifiers across four different dataset conditions. Models trained and tested on traditional synthetic traffic—collected via scripted Puppeteer crawlers—achieve up to 98\% accuracy, consistent with prior literature. However, when these same models are tested on traffic generated by real users interacting with dynamic websites, accuracy drops sharply to below 10\%, even for the strongest architectures. This illustrates that existing synthetic datasets fail to capture the interaction complexity necessary for evaluating WFP in modern contexts.

Even within the real human dataset, classifier performance varies greatly depending on how the training data is partitioned. When we train on multiple participants and test on a held-out individual (leave-one-user-out), accuracy declines by over 30\%, even under closed-world assumptions. This gap demonstrates the importance of modeling user-specific behavior and highlights that dataset diversity—not just scale—is essential for building robust WFP systems.

Collecting real user traffic to meet these requirements is expensive. Our experiments show that acquiring 1GB of authentic human browsing data costs approximately \$35, primarily due to participant compensation and lab infrastructure. In comparison, our LLM-based multi-agent framework can generate 1GB of synthetic—but human-like—traffic for just \$10 using commercial APIs. We anticipate this cost gap will widen further as open-source LLMs and lower compute costs continue making high-quality synthetic data generation more feasible.

\subsection{Key Observations and Motivation for Multi-Agent LLM-Based Generation}

In re-evaluating Website Fingerprinting (WFP) attacks under modern conditions using real user traces and stricter constraints (as formalized in Section~\ref{sec:threatmodel}), we identified three critical challenges that limit the generalizability of current approaches:

\textbf{Observation 1: WFP models suffer significant accuracy degradation under realistic conditions.}  
Even in a closed-world setting, model accuracy dropped from previously reported 95\%+ to around 80\% when tested on unsegmented, interaction-rich traffic segments—highlighting the brittleness of prior assumptions such as clean session boundaries.

\textbf{Observation 2: Model performance is highly sensitive to dataset scale and diversity.}  
Modern websites span heterogeneous services, components, and user flows. We find that classification performance improves significantly with larger and more varied training sets, even within the same domain—underscoring the need for behavioral diversity.

\textbf{Observation 3: User-specific behaviors hinder generalization.}  
When evaluating on users excluded from the training set, model accuracy dropped sharply. This indicates that models often overfit to individual browsing styles, failing to extract site-invariant features.

These limitations collectively reveal a central bottleneck: achieving robust WFP in modern environments requires access to large-scale, semantically rich, and behaviorally diverse datasets. Collecting such traffic from real users is both costly and ethically fraught, due to privacy risks and limited scalability.

To address this, we introduce a scalable multi-agent data generation pipeline powered by Large Language Models (LLMs). By prompting commercial LLMs such as Claude with user personas and site-specific goals, and executing their outputs through a structured browser control API, we synthesize realistic traffic patterns that closely mirror human browsing behavior. Our approach balances semantic fidelity and controllability, enabling fine-grained simulation of diverse user interactions at a low cost. As we demonstrate in later sections, LLM-generated data not only improves generalization to unseen users but also outperforms traditional synthetic traffic by a wide margin, providing a viable and scalable alternative for future WFP research and evaluation.

\section{Threat Model}
\label{sec:threatmodel}

We consider a passive network-level attacker attempting to identify the websites visited by a user based solely on encrypted traffic metadata. This threat model is rooted in traditional WFP literature~\cite{panchenko2011website, hayes2016kfingerprinting}, but we adopt a more constrained and realistic version, reflecting what a modern adversary could plausibly observe.

\subsection{Attacker Capabilities}

The attacker monitors network flows from a position such as an ISP, Tor entry node, or VPN exit. They have access to:
\begin{itemize}
    \item Packet timestamp, size, and direction (in/outbound),
    \item Encrypted transport-layer traffic,
    \item Long-lived connections without application-layer insight.
\end{itemize}

Critically, unlike most prior work, the attacker in our setting does \textbf{not} observe:
\begin{itemize}
    \item Session start or end events,
    \item Page load transitions or URLs,
    \item Any browser-level metadata or DOM structure.
\end{itemize}

The attacker must operate on \textit{arbitrary continuous segments} of encrypted packet traces, which may begin or end at any point within a browsing session. This assumption mirrors practical conditions where traffic segmentation is unavailable or noisy.

\subsection{Attacker Goal}

Given a packet segment $T$, the attacker's classifier $f(\cdot)$ will predict the corresponding label $\hat{w} \in W$, where $W$ is a closed set of known monitored websites. The segment may reflect part of a page load, a continuous user interaction (e.g., scrolling), or even a mixture of unrelated background requests.

\subsection{Differences from Prior Models}
% \hao{cite some prior work here?}
Our threat model introduces several key differences compared to prior work~\cite{online_website,df,deng2024robustreliableearlystagewebsite,tf,bapm,CHEN2021108298}:

\begin{itemize}
    \item \textbf{No clean session boundaries:} Realistic traces are unsegmented and may contain noise or overlap.
    \item \textbf{User behavior variability:} Per-user differences in interaction style, device usage, and attention span lead to substantial intra-class diversity.
    \item \textbf{Modern dynamic content:} Websites dynamically change in response to scrolling, login state, personalization, etc., increasing non-determinism in packet flows.
\end{itemize}

These constraints substantially raise the bar for WFP attacks, forcing classifiers to generalize beyond static features and idealized trace layouts. As we show in later sections, models trained under traditional assumptions significantly underperform in this setting—unless provided with large, diverse, and realistic training data.

Moreover, our threat model is designed to better reflect practical attacker conditions where ideal segmentation and deterministic traffic no longer hold. Prior defenses such as Walkie-Talkie~\cite{wang2017walkie}, TrafficSliver~\cite{trafficSliver2021}, and ZeroDelay~\cite{gong2020zero} rely on synchronous traffic shaping or burst modeling under the assumption of clean page boundaries and uniform session alignment. However, our model operates in a streaming scenario where segments are untethered from specific user actions, limiting the effectiveness of such defenses.

Additionally, our setting aligns with recent critiques of traditional evaluation assumptions~\cite{juarez2016critical, sirinam2018systematic, ohnishi2023subverting}, and is intentionally constructed to reveal the fragility of models trained on idealized data. By adopting this more realistic threat model, we demonstrate that modern defenses should be reassessed under unconstrained, long-lived browsing sessions that better match real-world conditions.

\section{Methodology}
\label{sec:methodology}

To rigorously evaluate website fingerprinting (WFP) under modern browsing conditions, we designed and collected a suite of traffic datasets\footnote{\url{https://drive.google.com/file/d/1KZNMZSUdLLBA5hQ0KNa165eI5zOADnkf/view?usp=sharing}} from three distinct sources: (1) real human users interacting with websites in a controlled lab environment, (2) scripted browsers that follow predefined sequences, and (3) a novel LLM-based multi-agent system capable of generating diverse and semantically meaningful interactions. Each source introduces different tradeoffs in realism, cost, and scalability.

This section presents the full methodology behind our data collection pipeline. We begin by describing our human participant study, including recruitment, task design, and privacy safeguards. We then detail the construction of a high-fidelity Puppeteer-based crawler for baseline automation. Finally, we introduce our LLM-driven synthetic framework with reward-based prompt optimization, which balances behavior diversity and command accuracy at scale.

\begin{figure}[t]
    \centering
    \includegraphics[width=\linewidth]{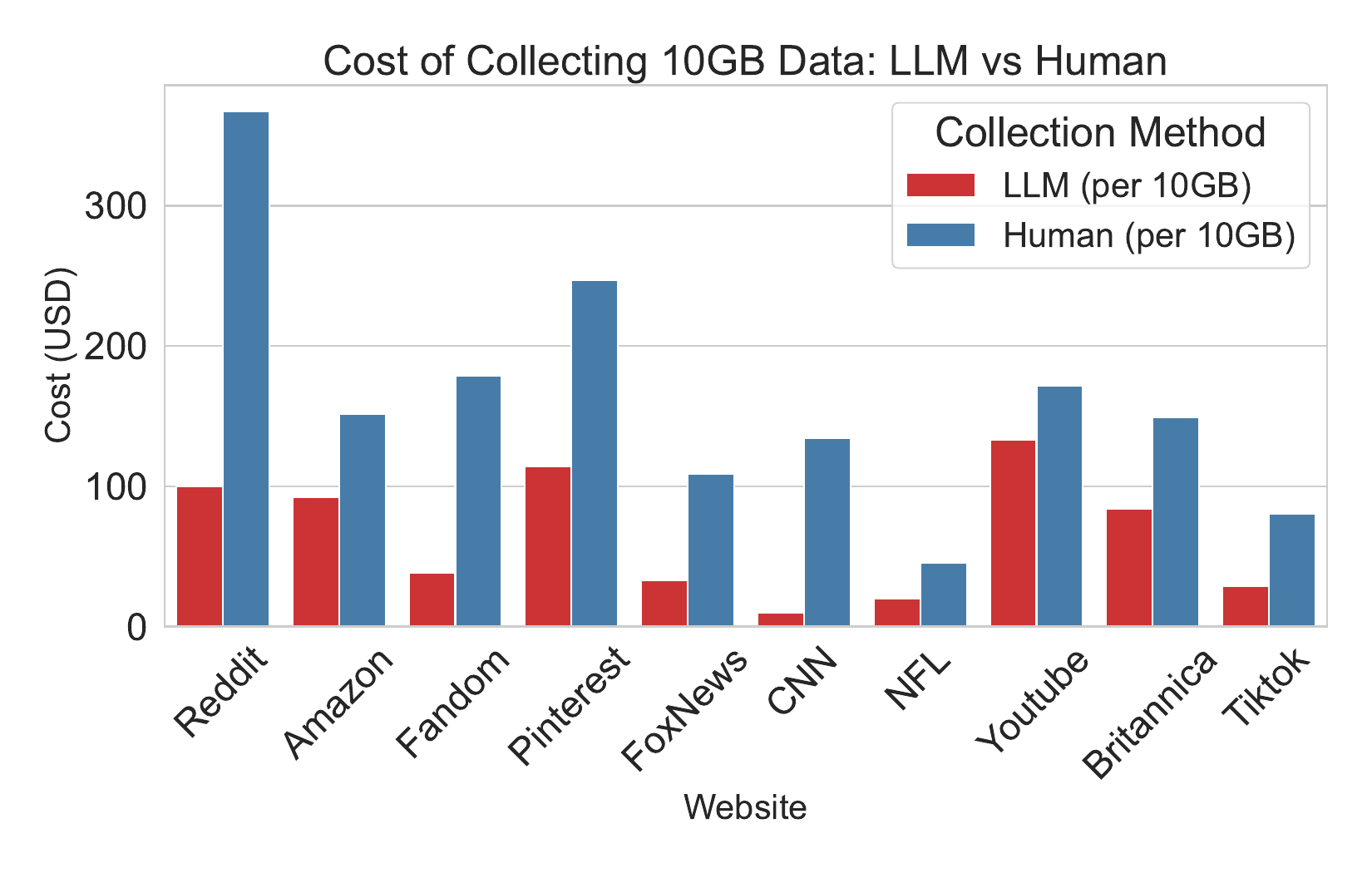}
    \caption{
    Cost comparison for collecting 10GB of web browsing data across ten websites using human participants versus LLM-based agents. LLM simulation offers up to 3–5$\times$ cost reduction while maintaining site-specific coverage.
    }
    \label{fig:llm-vs-human-bar}
\end{figure}
\begin{figure}[t]
    \centering
    \includegraphics[width=\linewidth]{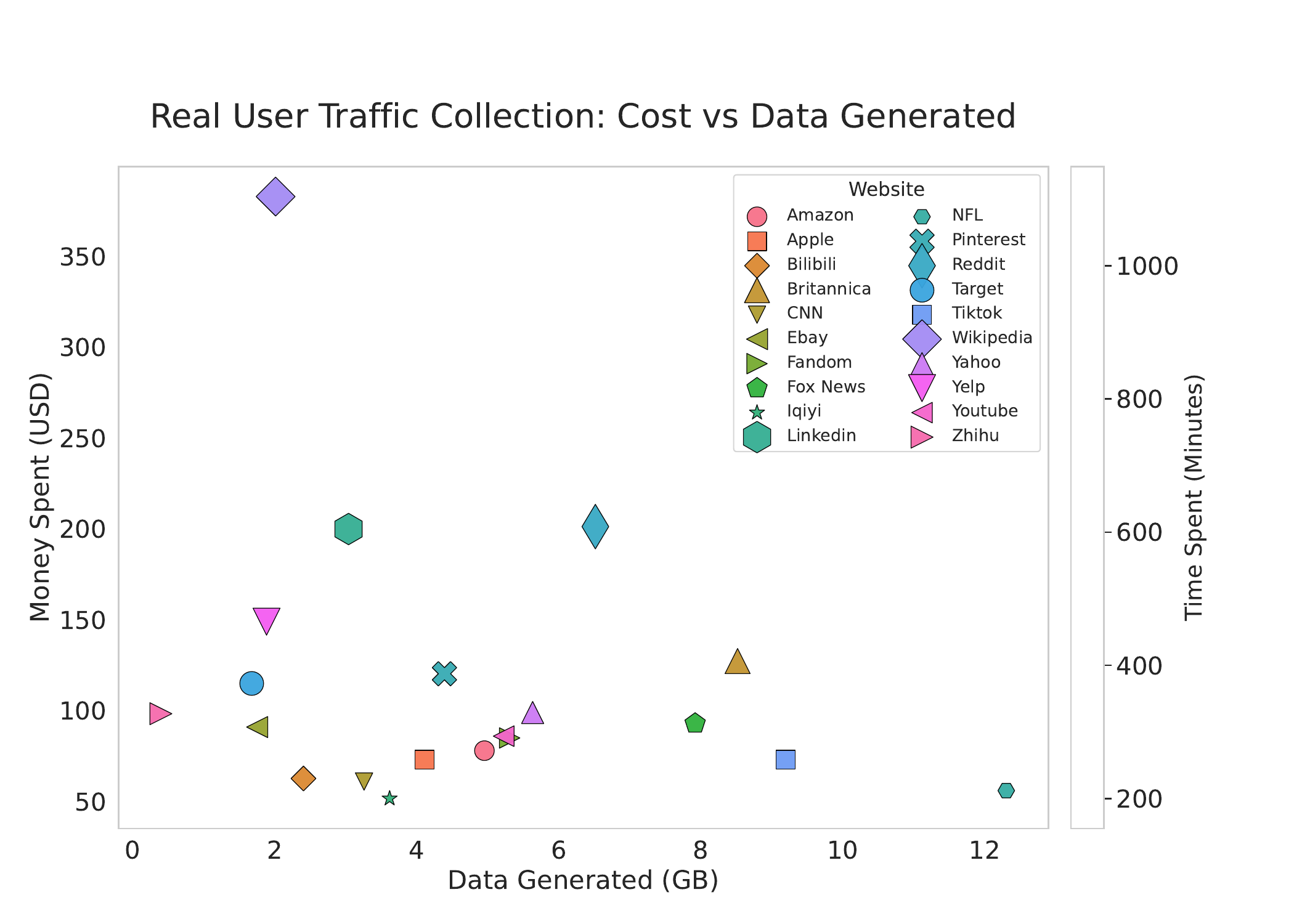}
    \caption{
    Traffic collection cost and yield per website using real participants. Bubble size and color reflect time spent per site. Some websites (e.g., \textit{Wikipedia}, \textit{LinkedIn}) incurred high cost but produced relatively little data, underscoring the inefficiency of manual collection at scale.
    }
    \label{fig:real-human-bubble}
\end{figure}

\subsection{Real Human Browsing Data Collection}

To construct a realistic dataset that captures genuine human interaction patterns on modern websites, we recruited and compensated 30 participants to browse 20 popular websites in a controlled laboratory setting. Each participant was paid \$20 per hour, and collectively, the study yielded over 100 hours of labeled browsing activity. Across all sessions, we gathered a total of \textbf{96.1 GB} of packet-level network traces.

The selected websites span a broad range of categories—including e-commerce (e.g., Amazon, eBay, Target), social media and video platforms (e.g., Reddit, YouTube, TikTok), and information-rich portals (e.g., CNN, Wikipedia, Britannica). A full list of target sites and task structures is included in section~\ref{sec:dataset}.

Before participation, each subject completed a demographic and behavioral background questionnaire capturing attributes such as age, gender, native language, and general browsing habits. Participants also rated their familiarity with each target website using a 5-point Likert scale. These profiles enabled us to analyze how prior knowledge and intent shape browsing behavior across different user types.

Each participant was assigned an isolated desktop workstation on a university-managed wireless network. They were instructed to use a standard Chromium browser without any extensions or personalization. For each website, a small set of mandatory tasks was provided to ensure consistent functional coverage (e.g., “search for a product” or “read a top article”), after which users were free to explore the site without restriction. Most sessions per website lasted between 20 and 40 minutes.

All network traffic was captured using \texttt{tcpdump}, generating full-fidelity packet traces in \texttt{pcap} format. These traces record packet size, direction, and timestamp—but no application-layer content—making them ideal for website fingerprinting analysis while preserving user privacy.

As shown in Figure~\ref{fig:real-human-bubble}, we observed substantial variance in data yield and collection cost across websites. For example, sessions on Wikipedia or LinkedIn often required over an hour of participant time but produced less than 3 GB of data, whereas NFL and Fandom yielded high-volume flows with relatively little user effort. This heterogeneity reflects differences in page architecture, media density, and engagement dynamics. The average cost per gigabyte across all 20 websites was approximately \$35, with certain domains exceeding \$100/GB—highlighting the scalability and economic limits of manual dataset creation.

\subsubsection{Ethical Considerations and Privacy Protections}
% log into personal accounts,
Given that our study involved human subjects and generated behavioral network traces, we took comprehensive steps to ensure ethical compliance and participant privacy. Our protocol was reviewed and approved by the Institutional Review Board (IRB) of our university. All participants gave their informed consent prior to the study and were explicitly instructed not to submit personal information or interact with external websites during the sessions.

To protect participant anonymity, we designed a hash-based identifier scheme that decouples recorded traffic from any real identity. Each session was labeled using a salted SHA-256 hash of an anonymized participant code and session timestamp. This identifier allows per-session trace linkage for analysis, without enabling user re-identification.

All traffic traces were encrypted using AES-256 and securely uploaded to a university-managed research cloud storage system with strict access control. Metadata such as website domain, session duration, and task type were stored separately using the same hash ID, allowing contextual analysis without a direct identity link. All questionnaire responses were aggregated, and any direct or indirect personal identifiers were either removed or anonymized before conducting the final analyses.

\subsection{Scripted Browsers Traffic Collection}
\sloppy
To establish a programmatic baseline for browsing behavior, we implemented a scripted crawler following the structure of SimulatedHuman~\cite{wfnet}. This crawler was designed to automate common surface-level actions across websites using hand-coded interaction flows. While it introduces basic variability—such as random delays, selecting links, or toggling tabs—its behavior is ultimately limited to deterministic scripts and lacks the adaptability seen in human or LLM-based browsing.

We built the crawler using \texttt{puppeteer-extra}, a headless Chromium automation framework equipped with stealth plugins to avoid bot detection. For each target site, we defined a set of site-specific scripts that execute static sequences: e.g., selecting a product, playing a video, or navigating to the next page. Some scripts simulate user delays or random branching, but none dynamically adapt based on page content or state changes. Compared to LLM agents, which make context-aware decisions based on semantic goals and visual state, scripted crawlers operate with shallow logic and do not adjust to personalized content or asynchronous interface events.

All sessions were run under the same devices and network conditions used in our human experiments to ensure a fair comparison. Over multiple days, we collected over \textbf{800 GB} of packet-level traces from 20 major websites using \texttt{tcpdump}. While this approach improves over classical crawlers in realism and scale, it remains inherently constrained by static control logic. As shown in later sections, models trained on this data fail to generalize to real user behavior, reinforcing the need for deeper, semantically grounded simulation frameworks.

\subsection{LLM Multi-Agent Data Generation with Online Prompt Optimization}

Deploying and training a fully self-hosted multi-agent LLM system\footnote{\url{https://drive.google.com/file/d/1Wf8X2Ncp80wbAuLm2W7GNt1tgh4jnN4f/view?usp=sharing}} with multi-modal input and online feedback would typically require significant GPU resources and infrastructure. To circumvent this, we designed a lightweight but practical training and deployment framework built on top of commercial Large Language Model (LLM) APIs—in our case, the Claude API. This allowed us to integrate high-quality language understanding and generation capabilities into a multi-agent system without the need to fine-tune a full LLM model.

% The system consists of two collaborative agents: a \textit{Decision-Making Agent} responsible for high-level reasoning and goal formulation, and a \textit{Computer-Using Agent} (CUA) that executes commands in a real browser via Puppeteer.

\paragraph{Multi-Agent System Configuration.}
Our multi-agent system is composed of two modular LLM-driven components tailored for semantic goal planning and browser-level execution. The \textbf{Decision-Making Agent} is powered by Claude 3.7 Sonnet—a multimodal model capable of interpreting both structured text and visual inputs. At each decision point, it analyzes the current page content and outputs a concise instruction grounded in what is visible. To enhance determinism and reduce hallucination, we fix the generation temperature to 0.3, ensuring stable command quality.

The \textbf{Computer-Using Agent (CUA)} runs inside a lightweight Docker container using Anthropic’s official execution environment. It interacts with a live Chromium browser by parsing the DOM and autonomously executing fine-grained GUI actions such as scrolling, clicking, and typing. This Docker-based setup provides secure isolation and simplifies deployment across machines, while also enabling scalability and multi-browser extensibility in future extensions.

Together, these agents operate in a closed-loop fashion: one agent formulates high-level goals, while the other translates them into executable actions. This architecture decouples semantic planning from browser control, offering robustness and transferability across websites and user roles.

During each interaction cycle, the Decision Agent receives four inputs:
\begin{enumerate}
    \item A prompt template encoding a persona and goal;
    \item The name of the target website;
    \item A full-page screenshot in base64 format;
    \item The role or user context (e.g., first-time user, returning visitor).
\end{enumerate}

% The system consists of two collaborative agents: a \textit{Decision-Making Agent} responsible for high-level reasoning and goal formulation, and a \textit{Computer-Using Agent} (CUA) that executes commands in a real browser via Puppeteer. The CUA operates inside a Docker container and interfaces with an instrumented Chromium instance, simulating realistic browsing actions like clicking, typing, or scrolling.

% During each interaction cycle, the Decision Agent receives four inputs:
% \begin{enumerate}
%     \item A prompt template encoding a persona and goal;
%     \item The name of the target website;
%     \item A full-page screenshot in base64 format;
%     \item The role or user context (e.g., first-time user, returning visitor).
% \end{enumerate}

Based on these, the LLM (Claude) returns a concise, executable natural language command. The command is then interpreted by the CUA and performed in the browser. After execution, success/failure is recorded along with whether the action was redundant or led to a crash. All interactions are logged to a structured JSONL file with command, reward, prompt ID, and screenshot hash.

\paragraph{Prompt Design Considerations.}
A central challenge in our multi-agent pipeline lies in ensuring that the high-level instructions produced by the Decision Agent are both \textit{semantically valid} and \textit{executable} by the downstream CUA. Large Language Models (LLMs), particularly when acting as generative planners, are prone to verbosity, hallucination, and nondeterministic outputs—especially under high sampling temperatures or vague prompting. In our context, these issues translate into commands that may be overly long, ambiguous, or mismatched with the visual state observed by the CUA.

For instance, an under-constrained Decision Agent might generate speculative, multi-sentence descriptions rather than screen-grounded commands (e.g., “You should probably compare different headphone brands and read some reviews first”), making it difficult for the CUA to interpret and execute the intended action. Worse, it may hallucinate interface elements that are not currently visible on the page, resulting in execution crashes.

To address this, we design instruction prompts that explicitly constrain generation to a single, executable, and screen-aligned command. Typical prompt examples include: “Generate \textbf{one realistic command to execute} based strictly on what is currently visible in the browser,” or “\textbf{Avoid explanation}; only output the next user action.” These constraints improve stability, filter verbosity, and increase the alignment between the Decision Agent’s intent and the CUA’s capabilities. Moreover, by refining these prompts through continual reward-based feedback, we are able to maintain high interaction fidelity and consistency over long multi-step trajectories.

\paragraph{Online Prompt Optimization.}
The quality of generated commands---especially their validity and diversity---directly affects the resulting synthetic dataset. To improve this, we apply an online prompt selection and evolution framework based on a classical multi-armed bandit formulation. Each prompt template $p_k$ is treated as a separate arm, and its expected reward $\mu_k$ is learned online through interaction. We use an $\epsilon$-greedy policy for balancing exploration and exploitation: with probability $\epsilon$ a prompt is randomly sampled; otherwise, the one with the highest estimated reward is chosen.

\paragraph{Continuous Reward Formulation.}
Let $s_t \in \{0, 1\}$ indicate whether the command executed successfully, and let $c_t \in \mathbb{N}$ be the number of polling rounds required to complete execution. We define the continuous crash penalty $k_t$ based on execution time as:
\begin{equation}
    k_t = \min\left(1, \frac{c_t}{C_{\text{max}}} \right),
\end{equation}
where $C_{\text{max}}$ is the maximum allowed polling steps (e.g., 120). Let $d_t \in [0, 0.2]$ denote a diversity bonus computed from the dissimilarity between the current command and recent command history.
The total reward is then given by:
\begin{equation}
    R_t = s_t + d_t - k_t,
\end{equation}
which smoothly penalizes longer execution durations and rewards novel and successful commands.
The average reward for each prompt is updated online:
\begin{equation}
    \mu_k \leftarrow \mu_k + \frac{1}{n_k}(R_t - \mu_k),
\end{equation}
where $n_k$ is the number of times $p_k$ has been selected.
To prevent convergence to a narrow set of overfit prompts, we implement a prompt evolution mechanism: every $T$ rounds (e.g., $T = 10$), we prune the bottom 20\% of low-reward prompts and replace them with new rewrites generated by Claude. This allows the prompt pool to continuously adapt and diversify over time.

\vspace{0.5em}
%% \begin{algorithm}[H]  --had to comment the [H]  out to compiler locally --RPM
\begin{algorithm}  
\caption{Prompt Optimization via Multi-Armed Bandit (Extended Reward)}
\KwIn{Prompt pool $\mathcal{P} = \{p_1, \dots, p_K\}$, $\epsilon$ (exploration rate)}
\KwOut{Updated reward values $\mu_k$ for all prompts}
\ForEach{round $t = 1$ to $T$}{
    \tcp{Select a prompt}
    With probability $\epsilon$, randomly select $p_k \in \mathcal{P}$\;
    Otherwise, select $p_k = \arg\max_{p} \mu_p$\;

    \tcp{Construct input and call LLM}
    Instantiate $p_k$ with user context and target site\;
    Encode current screenshot as base64 image $I$\;
    Send $(p_k, I)$ to Claude API to get command $c$\;

    \tcp{Execute command}
    Let $(s_t, c_t) \gets$ \texttt{executeCommand}$(c)$\; \tcp{$s_t$: success, $c_t$: check count}
    Compute $k_t = \min(1, c_t / C_{\text{max}})$\;
    Compute $d_t \gets$ diversity score w.r.t. command history\;
    Compute $R_t = s_t + d_t - k_t$\;

    \tcp{Update reward statistics}
    $n_k \gets n_k + 1$\;
    $\mu_k \gets \mu_k + \frac{1}{n_k}(R_t - \mu_k)$\;

    \tcp{Log execution record}
    Save $\langle p_k, c, R_t, s_t, k_t, d_t \rangle$ to JSONL log\;

    \If{$t \mod 10 = 0$}{
        \tcp{Prune and replace prompts}
        Remove bottom 20\% prompts from $\mathcal{P}$\;
        Replace with new prompts rewritten by Claude\;
    }
}
\Return{$\{\mu_k\}_{k=1}^K$}
\end{algorithm}
\vspace{0.5em}

This reward-driven optimization loop allows our multi-agent framework to produce longer, more diverse interaction trajectories across multiple websites---resulting in synthetic traffic traces that better approximate the distributional properties of real human users.

Beyond diversity and realism, another key advantage of our approach lies in its scalability and cost-efficiency. Compared to collecting traffic from real human participants, our LLM-based multi-agent framework offers substantial cost reductions while maintaining semantic fidelity. While exploring all cost factors involved in collecting training traffic is beyond the scope of this paper, Figure~\ref{fig:llm-vs-human-bar} allows us to compare the marginal costs of humans to an LLM approach. We found that generating 10GB of browsing data using human participants can cost between \$80 and \$300 per site due to participant compensation, experimental supervision, and limited parallelism. In contrast, our framework—built on top of commercial LLM APIs—can synthesize comparable volumes of traffic at a fraction of the cost (typically under \$30 per 10GB) while scaling seamlessly across machines and websites. We were unable to compare the costs of obtaining traffic using scripted browsers as we used existing University servers and networking bandwidth free of charge. A more comprehensive analysis might look at cloud computing charges or the costs of buying or renting servers. 

The LLM agents require active prompt optimization that ensures high behavioral variance and realistic interaction patterns. By integrating reward-based self-improvement and structured feedback, the system maintains both command precision and browsing diversity over long sessions. This enables a synthetically generated WFP traffic corpus to serve as a viable complement, and, in many settings, an alternative, to costly human-collected data.

\section{Evaluation}
\label{sec:experiments}
We comprehensively evaluate the generalization performance of modern website fingerprinting (WFP) models under realistic web conditions. Our experiments are structured in four stages: we first introduce our experimental setup and curated datasets, then assess model performance under traditional versus real-user traffic (Stage~I). We next explore whether large language model (LLM)-driven multi-agent traffic can enhance cross-user generalization (Stage~II), and finally analyze how training sample size influences robustness across scripted and LLM-generated data regimes. Together, these evaluations reveal the limitations of legacy Scripted Browsing traces, highlight the value of behaviorally rich simulations, and motivate new scalable methodologies for WFP research.
\subsection{Experimental Setup}

To evaluate the resilience and generalization of modern website fingerprinting models, we benchmark nine state-of-the-art classifiers across a wide range of datasets and training conditions. These models encompass convolutional neural networks (CNNs), transformers, contrastive learning methods, and specialized fingerprinting architectures.
% Each was re-implemented or adopted from open-source repositories using author-recommended hyperparameters and input representations.
\begin{itemize}
    \item \textbf{TMWF}~\cite{tmwf} employs a temporal mutual information framework to align features extracted from raw packet sequences. By maximizing temporal consistency, it captures both short-term bursts and long-range temporal correlations for accurate classification.
    
    \item \textbf{ARES}~\cite{ares} proposes a robust, tab-agnostic encoder that disentangles site identity from tab-related noise, enabling generalizable fingerprinting across complex multi-tab sessions. It operates without requiring session segmentation or tab labels, enhancing real-world applicability.
    
    \item \textbf{NetCLR}~\cite{netclr} applies self-supervised contrastive learning on unlabeled packet sequences to learn discriminative representations. Its dual-branch architecture uses augmentations in both the temporal and directional space to enforce alignment and variance in latent embeddings.
    
    \item \textbf{BAPM}~\cite{bapm} introduces a block attention mechanism tailored for multi-tab browsing scenarios, effectively handling overlapping traffic by segmenting traces into distinct blocks. Unlike prior models, it fully utilizes the entire packet trace, including overlapping areas, to enhance website fingerprinting accuracy. 
    
    \item \textbf{TF}~\cite{tf} models WFP as a metric learning problem using triplet networks, enabling few-shot learning with only a handful of samples per class. It offers strong generalization with minimal training data.
    
    \item \textbf{Var-CNN}~\cite{varcnn} leverages a variable-kernel CNN with residual connections and dilated convolutions. By processing packet sizes and directions jointly, it provides robustness to timing noise and user behavior variation.
    
    \item \textbf{Tik-Tok}~\cite{tiktok} proposes a traffic splitting mechanism that segments flows into semantically meaningful bursts (or “tiks” and “toks”). Each burst is classified individually, and predictions are aggregated over time to adapt to dynamic web content loading.
    
    \item \textbf{DF}~\cite{df} is one of the most widely adopted WFP baselines. It combines deep CNN layers with dropout and batch normalization, using direction-augmented input matrices. DF remains a strong baseline due to its simplicity and surprising effectiveness across datasets.
    
    \item \textbf{WFNet}~\cite{wfnet} integrates a pre-trained traffic encoder with a fine-tunable classification head. It is trained using a domain-aware transfer learning pipeline, making it especially effective under low-resource or cross-domain conditions.
\end{itemize}

All models are trained and evaluated under a consistent pipeline using PyTorch on a shared GPU cluster. The input format follows the convention of 5000-length packet size/times sequences. That is, a \textbf{sample} is a 5000 length vector of packet sizes and times from a contiguous block of 5000 packets. Packet traces were divided into sequences of non-overlapping blocks to generate samples. Each experiment uses the same closed-world label space (e.g., 5, 10, or 20 websites, depending on the setting), and performance is averaged across three random seeds to mitigate variance.

\subsection{Dataset Overview: Realistic Web Fingerprinting at Scale}
\label{sec:dataset}

To evaluate the effectiveness of WFP models under realistic web conditions, we curate and utilize three distinct datasets, each designed to capture varying degrees of interaction realism and user diversity. These datasets include: (1) a large-scale real human browsing dataset, (2) a scripted browsers dataset, and (3) an LLM-simulated interaction dataset generated through our multi-agent framework.

\paragraph{RealHuman Dataset.} 
Our primary dataset consists of real user traffic collected from 20 modern websites, chosen to reflect the complexity and diversity of contemporary web usage. These sites span e-commerce (\textit{Amazon, Ebay, Target}), media and news (\textit{CNN, Fox News, Britannica, NFL}), social platforms (\textit{Reddit, Pinterest, Fandom, TikTok, Zhihu}), streaming services (\textit{YouTube, Bilibili, Iqiyi}), and utility portals (\textit{LinkedIn, Yahoo, Yelp, Apple, Wikipedia}). The site selection is inspired by recent research such as WFNet~\cite{wfnet}, emphasizing dynamic and multi-service platforms. Many of these sites embody modern Single-Page Application (SPA) architectures, where rich content is dynamically loaded and user interaction spans multiple services. For instance, a user session on Amazon may include product search, recommendation browsing, video playback, and cart operations---each generating different traffic patterns without discrete page boundaries.

We hired real-world participants to browse these websites freely across a range of devices and conditions. Each participant was instructed to engage naturally with the content. After over 100 hours of browsing, we collected \textbf{800 samples} per class, approximately equal to \textbf{4.5 GB of traffic per website}. This dataset provides a realistic, behavior-rich benchmark for evaluating WFP robustness.
% totaling \textbf{1000 training samples per class}

\paragraph{Scripted Browsers Dataset.} 
To enable comparison with prior work, we also generate a large-scale scripted browsing dataset using a scripted crawler. This dataset mimics conventional WFP collection protocols where a bot sequentially visits predefined URLs. The crawler does not emulate user interaction such as scrolling, pausing, or branching behavior. It simply loads a page and proceeds after a fixed timeout.

While this strategy supports efficient data collection, it fails to capture the semantic and behavioral diversity inherent in real human traffic. Nonetheless, we produce \textbf{8000 samples} per class, corresponding to approximately \textbf{20 GB of traffic}, for controlled benchmarking. This allows direct comparison with historical WFP results and highlights the limitations of purely rule-based sampling in the context of modern, interactive sites.

\paragraph{LLM-Sim Dataset.}
Finally, we introduce a third dataset created through our LLM-based multi-agent browsing framework. Unlike traditional crawlers, these agents follow natural-language browsing goals and simulate semantically grounded interaction sequences across websites. For computational efficiency and API constraints, we focus on a 10-site subset from our RealHuman corpus, including \textit{Amazon, Britannica, CNN, Fandom, Fox News, NFL, Pinterest, Reddit, TikTok, YouTube}. Each site is represented by \textbf{1000 samples}, yielding approximately \textbf{7 GB of traffic per website} and totaling 10,000 traces overall.

The reduced dataset size is a result of rate limits imposed by commercial Computer Use APIs and LLM-hosted browser environments. Despite the lower volume, our results demonstrate that LLM-generated data---though smaller---can surpass script-based baselines in generalization performance (see Section~\ref{sec:StageII}). This underscores the efficiency of behaviorally diverse data and the promise of LLM agents as realistic traffic generators.

\paragraph{Summary.} In total, our evaluation spans over 180,000 browsing traces across three datasets and three interaction modes (scripted, human, and simulated). These datasets allow us to isolate the effects of behavioral realism, semantic intent, and interaction complexity on WFP model performance under both traditional and adversarial training conditions.

\begin{figure}[t]
    \centering
    \includegraphics[width=\linewidth]{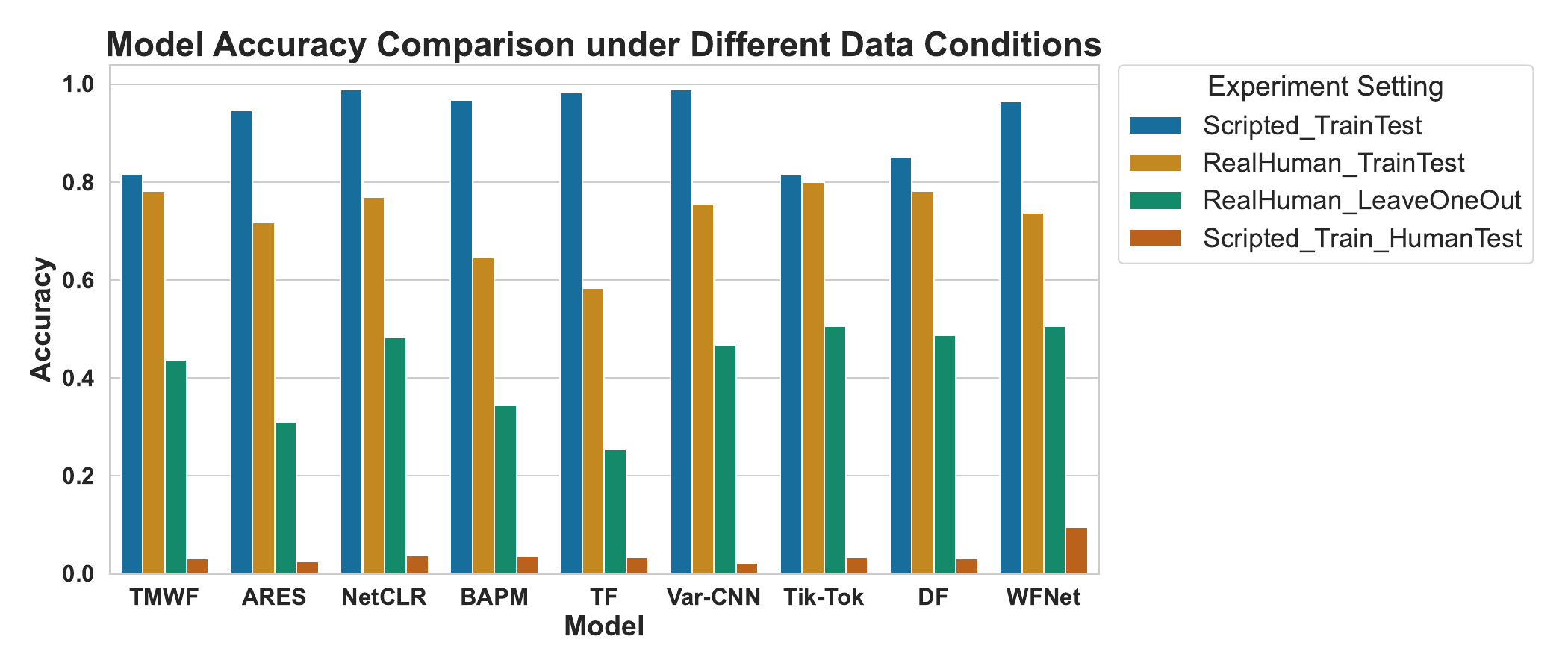}
    \caption{Accuracy of nine WFP models under four training/testing regimes: (1) \textsf{Scripted\_TrainTest}, (2) \textsf{RealHuman\_TrainTest}, (3) \textsf{LeaveOneUser}, and (4) \textsf{CrossDomain}. Results highlight the generalization gap between scripted and real browsing data.}
    \label{fig:model-accuracy-bar}
\end{figure}

\subsection{Stage I: Performance under Controlled and Realistic Conditions}
\label{sec:StageI}
\sloppy
We begin our evaluation by revisiting the standard assumption in the WFP literature: models are trained and tested under controlled, synthetic conditions using rule-based scripted crawlers. While these benchmarks have driven progress for nearly a decade, they often overlook the complexities of modern interactive websites. We compare model performance under two single-domain settings—robot-simulated traffic versus real human browsing data—before transitioning to cross-user and cross-domain generalization.

Our goal is twofold. First, we seek to quantify how interaction fidelity influences classifier performance: can WFP models still succeed when exposed to naturalistic, behavior-rich sessions? Second, we aim to establish a foundation for later sections by highlighting the limitations of scripted data and the need for richer behavioral diversity.

\paragraph{Single-Domain Evaluation: Scripted\_TrainTest vs. RealHuman\_TrainTest}

We evaluate nine state-of-the-art models under the \textsf{Scripted\_TrainTest} and \textsf{RealHuman\_TrainTest} conditions. As shown in Figure~\ref{fig:model-accuracy-bar} and Table~\ref{tab:metrics_summary}, models trained and tested on scripted traffic achieve near-perfect results (e.g., NetCLR: 98.9\%, TF: 98.3\%, Var-CNN: 98.9\%), confirming past findings that scripted data is trivially separable. This is largely due to the uniform click paths, predictable transitions, and stateless nature of robot-generated sessions, which fail to capture interaction variance.

However, accuracy drops significantly when models are trained and evaluated on real human traces. Even the strongest performers (TikTok, DF, NetCLR) plateau at 75--80\%, while others like TF and ARES drop below 65\%. F1-score, precision, and recall all decline, most notably recall, which reveals that models often miss diverse user behaviors. For instance, TF's F1-score falls from 0.983 to 0.583, but its recall drops even more steeply, suggesting its high-confidence predictions are limited to a few familiar patterns. In contrast, models like Var-CNN and NetCLR maintain higher recall (0.774 and 0.747 respectively), indicating stronger robustness to intra-site user variability.

This drop is not just a matter of accuracy but also consistency. Across our 20-site benchmark, human data introduces session noise, personalization artifacts, and diverse navigation flows. For example, different users may access entirely different parts of Amazon or YouTube depending on their intent, identity, and scroll depth. These complexities yield overlapping traffic signatures that challenge even state-of-the-art classifiers. The results underscore the limitations of relying solely on scripted browsing data and affirm the necessity of training on interaction-rich traces.

\paragraph{Generalization under LeaveOneUser and CrossDomain Tasks}

We next test model robustness under two increasingly difficult settings. In \textsf{LeaveOneUser}, models are trained on all but one human participant and tested on the held-out user. In \textsf{CrossDomain}, models are trained on scripted traffic and evaluated on real human traces.

The \textsf{LeaveOneUser} setting shows a consistent accuracy drop of 30--40\% across models (e.g., Var-CNN: 75.6\% to 46.8\%). Most models lose recall faster than precision, suggesting that they overfit to common behaviors while failing to generalize across unseen users. NetCLR and WFNet demonstrate relatively higher leave-one-out recall, indicating their resilience to behavioral diversity. Interestingly, WFNet also exhibits the highest recall in CrossDomain (0.102), reflecting a robust embedding mechanism that adapts well to unobserved interaction sequences.

\textsf{CrossDomain} is the most challenging task. Models trained solely on scripted traffic completely fail to generalize, with accuracy often falling below 10\% and F1-scores near zero. For instance, TF drops to 0.034, ARES to 0.025, and Var-CNN to 0.022 F1-score. These failures highlight the structural mismatch between scripted and real traffic--rule-based bots do not mimic personalized interaction.

These findings motivate our exploration of synthetic alternatives that better mirror human browsing. In Stage II, we assess whether LLM-based traffic can bridge this realism gap.

\begin{table*}[t]
\centering
\caption{Performance metrics (F1-score, Precision, Recall) of each model under four training/testing settings.}
\label{tab:metrics_summary}
\resizebox{\textwidth}{0.11\textheight}{
\begin{tabular}{l|ccc|ccc|ccc|ccc}
\toprule
\multirow{2}{*}{Model} 
& \multicolumn{3}{c|}{\textbf{Scripted\_TrainTest}} 
& \multicolumn{3}{c|}{\textbf{RealHuman\_TrainTest}} 
& \multicolumn{3}{c|}{\textbf{LeaveOneUser}} 
& \multicolumn{3}{c}{\textbf{CrossDomain}} \\
& F1 & Prec & Rec & F1 & Prec & Rec & F1 & Prec & Rec & F1 & Prec & Rec \\
\midrule
TMWF   & 0.816 & 0.855 & 0.818 & 0.782 & 0.803 & 0.775 & 0.437 & 0.489 & 0.489 & 0.031 & 0.024 & 0.084 \\
ARES   & 0.947 & 0.949 & 0.947 & 0.718 & 0.723 & 0.722 & 0.410 & 0.440 & 0.443 & 0.025 & 0.058 & 0.064 \\
NetCLR & 0.989 & 0.989 & 0.989 & 0.769 & 0.804 & 0.747 & 0.482 & 0.505 & 0.495 & 0.037 & 0.047 & 0.056 \\
BAPM   & 0.968 & 0.969 & 0.968 & 0.646 & 0.672 & 0.654 & 0.344 & 0.391 & 0.399 & 0.035 & 0.029 & 0.068 \\
TF     & 0.983 & 0.984 & 0.983 & 0.583 & 0.603 & 0.581 & 0.254 & 0.287 & 0.278 & 0.034 & 0.053 & 0.052 \\
Var-CNN& 0.989 & 0.989 & 0.989 & 0.756 & 0.760 & 0.774 & 0.468 & 0.528 & 0.512 & 0.022 & 0.019 & 0.051 \\
Tik-Tok& 0.815 & 0.859 & 0.831 & 0.800 & 0.812 & 0.801 & 0.505 & 0.538 & 0.532 & 0.035 & 0.043 & 0.044 \\
DF     & 0.852 & 0.910 & 0.845 & 0.782 & 0.794 & 0.785 & 0.487 & 0.534 & 0.511 & 0.031 & 0.041 & 0.069 \\
WFNet  & 0.965 & 0.965 & 0.965 & 0.737 & 0.763 & 0.740 & 0.495 & 0.525 & 0.532 & 0.096 & 0.091 & 0.102 \\
\bottomrule
\end{tabular}
}
\end{table*}

\begin{figure*}[t]
    \centering
    \includegraphics[width=\linewidth, height=0.26\textheight]{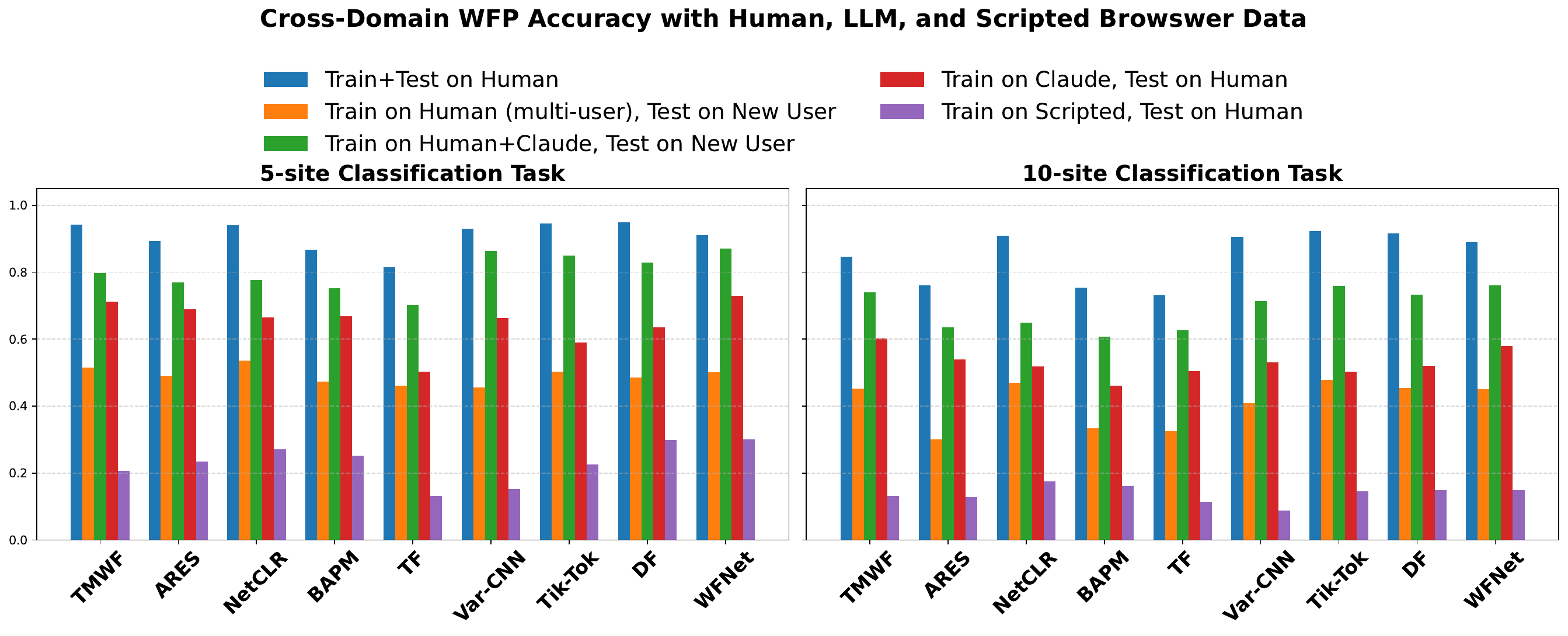}
    \caption{
    Accuracy comparison across five training strategies (Human only, Leave-One-User, Human+Claude, Claude-only, and Scripted-only) on two tasks: \textbf{Top:} 5-site classification, \textbf{Bottom:} 10-site classification. LLM-based traffic significantly boosts cross-user accuracy compared to synthetic baselines, despite using only 20\% of the data size.
    }
    \label{fig:llm-cross-transfer}
\end{figure*}

\subsection{Stage II: Cross-User Generalization Enhanced by LLM-Simulated Traffic}
\label{sec:StageII}
We now test whether synthetic traffic from our multi-agent LLM system can improve generalization. In this stage, we use the 5-site and 10-site variants of the closed-world task. As before, we focus on cross-user transfer: training on multiple users or simulations, and testing on a previously unseen human.

Before analyzing LLM-based enhancements, we briefly revisit the baseline human-only results in this reduced-domain setting. While \textsf{RealHuman\_TrainTest} offers reasonably strong performance due to random within-user splits, accuracy drops sharply under the \textsf{LeaveOneUser} condition. For instance, in the 10-site task, TikTok and NetCLR drop from over 80\% to roughly 50\%, and DF from 78.2\% to 48.7\%. This gap highlights the difficulty of modeling diverse human behavior across users. Without explicit exposure to intra-site variation in scrolling, clicking, or content consumption styles, models struggle to generalize—even when trained on large real-world datasets. These results reinforce the need for augmentation mechanisms that can synthetically reproduce such behavioral heterogeneity—precisely the goal of our multi-agent LLM framework.

\paragraph{Comparative Training Settings}

In addition to \textsf{Human\_TrainTest} and \textsf{LeaveOneUser} from Stage I, we evaluate:

\begin{enumerate}
    \item \textsf{Human+Claude}: Train on a combined set of human and LLM-generated traces.
    \item \textsf{ClaudeOnly}: Train on LLM-simulated traffic exclusively.
    \item \textsf{ScriptedOnly}: Train on scripted browers traces exclusively.
\end{enumerate}

\paragraph{Performance Gains and Explanation}

As Figure~\ref{fig:llm-cross-transfer} shows, \textsf{Human+Claude} consistently improves generalization. Accuracy jumps by 25--35\% in both tasks. For example, NetCLR improves from 53.6\% to 77.3\% in the 5-site task, and TikTok from 50.3\% to 84.9\%. Importantly, LLM data is particularly helpful for models like WFNet and BAPM, which benefit from longer temporal context and diverse semantic cues.

Models trained with \textsf{ClaudeOnly} also outperform \textsf{ScriptedOnly} baselines by large margins—often tripling the accuracy. Even with one-fifth the data, Claude simulations better approximate realistic behavior. This is because decision-making agents are instantiated with personas (e.g., “tech-savvy teen,” “price-sensitive shopper”) that guide behavior in context-sensitive ways. The instructions they generate capture richer, goal-driven browsing, which leads to more realistic interaction trajectories.

\paragraph{Behavioral Regularization and Robustness}

Claude agents help reduce inter-site variance and prevent overfitting to frequent paths. For example, TikTok and Var-CNN exhibit smoother performance across sites when trained with LLM data. The behavioral regularization effect arises from semantic grounding, consistent interaction patterns, and broader behavioral coverage—especially helpful for infinite-scroll or recommendation-based platforms.

Interestingly, in some cases LLM-trained models even outperform human-only training. For instance, WFNet scores 74.4\% on 10-site ClaudeOnly training versus 62.1\% on Human\_TrainTest. This suggests that simulated users can exceed real human diversity by covering edge-case paths or low-frequency interaction types.

\paragraph{Looking Ahead}

These results reinforce three core takeaways:

\begin{itemize}
    \item Traditional scripted traffic is insufficient for training WFP models on modern interactive websites.
    \item LLM-generated traffic introduces critical behavioral diversity, enhancing robustness to user variability.
    \item Multi-agent simulation offers a scalable alternative to expensive human data collection.
\end{itemize}

In the next section, we analyze how the number of LLM samples influences this benefit. We find that while increasing LLM data improves generalization, scaling scripted traffic yields minimal gains, further confirming the importance of realism over volume.

\subsection{Impact of Training Sample Size: LLM vs. Scripted Browsers Traffic}
\label{sec:sample-scaling}
\begin{figure}[t]
    \centering
    \includegraphics[width=\linewidth]{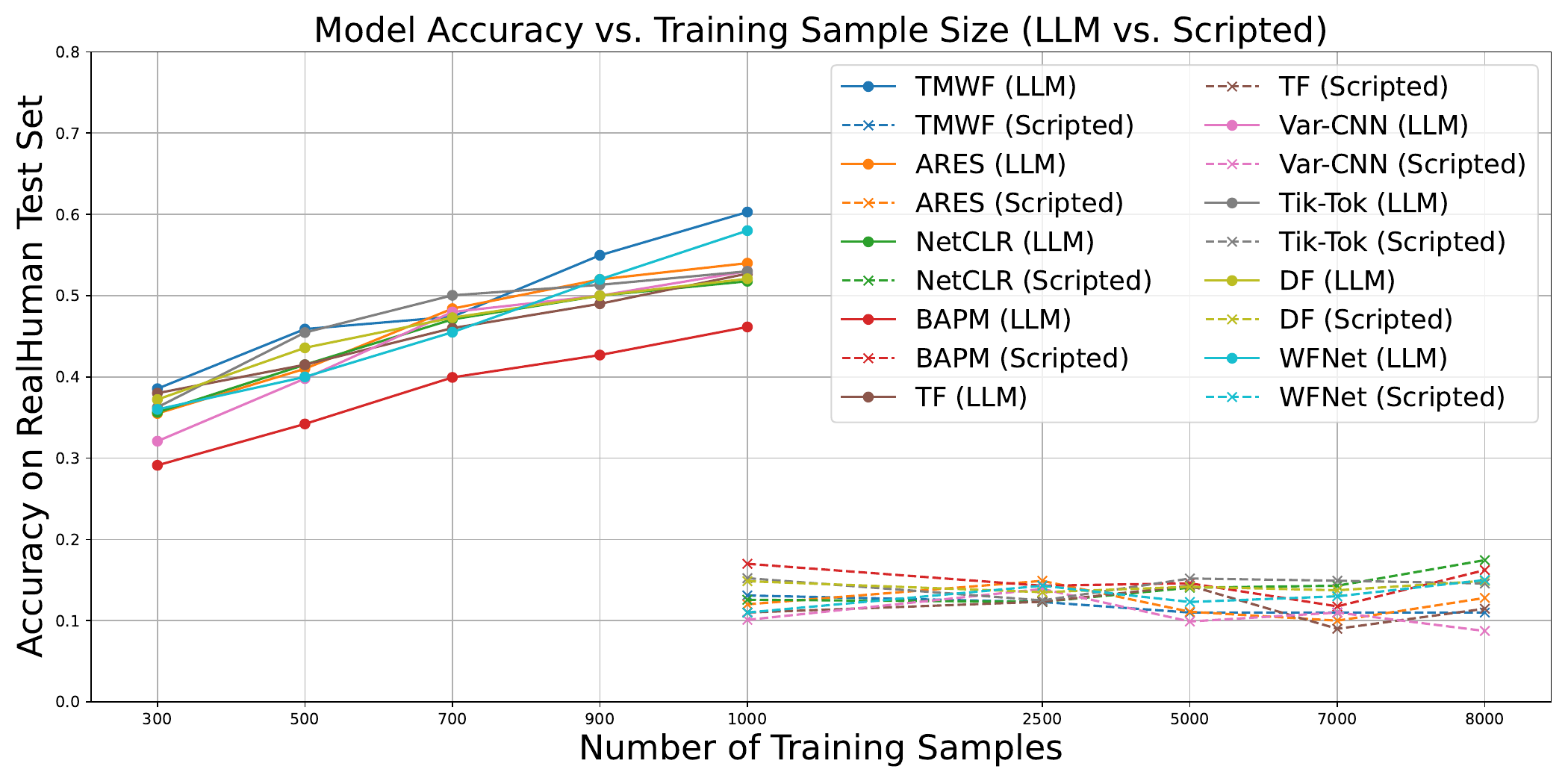}
    \caption{
    Accuracy of nine WFP models trained on LLM-generated (solid lines, 300--1000 samples) vs.\ scripted browser traffic (dashed lines, 1000--8000 samples). While scripted traffic offers little gain with scale, LLM data steadily improves generalization, highlighting that realistic interaction diversity, not volume alone, is key to modern WFP performance.
    }
    \label{fig:scaling-llm-synthetic}
\end{figure}

To further investigate how dataset scale influences generalization performance, we examine WFP model accuracy on a fixed \textsf{RealHuman\_Test} set as the number of training samples varies. Figure~\ref{fig:scaling-llm-synthetic} presents a comparative analysis of models trained on multi-agent LLM-generated traffic (300--1000 samples per class) and traditional scripted browser traffic (1000--8000 samples per class).

Across all nine models, LLM-based training exhibits a clear upward trend: accuracy improves consistently as more training data becomes available. For example, TMWF and NetCLR reach 60.3\% and 51.8\% accuracy respectively at 1000 LLM samples—substantially outperforming their performance at 300 samples, where accuracies drop below 40\%. This scaling behavior is consistent and monotonic, indicating that our LLM-based agents produce data distributions that grow meaningfully with size and maintain alignment with the behavioral complexity of real users.

In contrast, scripted browers data demonstrates no such benefit. Despite training on up to 8000 samples per class, most models fail to exceed 17\% accuracy. Moreover, many models (e.g., Var-CNN, TF, DF) show marginal or no improvement when the sample size increases from 1000 to 8000. This suggests a saturation effect—scripted traffic, originally designed for static and deterministic websites, lacks the semantic diversity and session-level richness necessary for generalization to dynamic modern environments.

These findings support our central claim: website fingerprinting on modern web platforms is a fundamentally more challenging problem than prior literature has assumed. Legacy traffic collection methods, tailored to page-based navigation and fixed site structures, cannot capture the behavioral entropy, personalized content loading, and dynamic UI reactivity that characterize contemporary web experiences. Simply scaling these scripted datasets does not resolve the generalization gap—more low-quality data does not translate to better performance.

In contrast, the LLM-simulated traffic we propose introduces semantically grounded, persona-driven interactions that better reflect user heterogeneity. Even with fewer samples, LLM data consistently enables stronger real-world generalization. These results underscore the necessity of modern, high-fidelity training sources for realistic WFP evaluation and pave the way for scalable, agent-based alternatives to manual user data collection.

\section{Related Work}
\label{sec:relatedwork}

\paragraph{Website Fingerprinting Models and Limitations}

Classic WFP research focused on closed-world classification using hand-crafted features and traditional ML models~\cite{panchenko2011website}. Deep Fingerprinting (DF)~\cite{df} introduced CNN-based models, followed by Var-CNN~\cite{varcnn}, Tik-Tok~\cite{tiktok}, and transformer-based methods like TF~\cite{tf} and WFNet~\cite{wfnet}, all of which improved closed-world accuracy under synthetic settings. Recent works (e.g., ARES~\cite{ares}, NetCLR~\cite{netclr}, TMWF~\cite{tmwf}) explored robustness through contrastive learning, domain transfer, and adversarial augmentation.

However, most prior studies rely on static websites and rule-based robot traffic, which oversimplify modern user behavior. Our work revisits this assumption and proposes evaluation under realistic conditions, highlighting the generalization gap in current WFP pipelines.

\paragraph{LLM-Based Agents and Human Interaction Simulation}

Recent advances in LLM-powered agents (e.g., CAMEL, Voyager, AutoGPT) have enabled simulated workflows involving planning, memory, and tool use~\cite{llm_survey}. Specifically, browser-interacting agents~\cite{sager2025aiagentscomputeruse, agentbench, ref:webarena,wshop} execute actions in real UIs using vision-language reasoning, but typically focus on task success, not traffic realism.

Our approach is the first to apply LLM-based agents to \textit{traffic generation} for WFP. By combining goal-driven LLM planning with real browser execution, we simulate diverse, high-entropy sessions that more faithfully reflect modern web usage. This bridges the realism gap and unlocks new directions for adversarial evaluation at the network level.

\section{Discussion}
\label{sec:discussion}

Despite the superior effectiveness of attacks trained on data generated by LLM-based multi-agent simulation compared to traditional scripted methods, our approach still faces important limitations and open challenges. In this section, we outline the tradeoffs, scalability concerns, and potential future research directions stemming from our findings.

\paragraph{Cost vs.\ Realism Tradeoffs in Data Generation}

Compared to traditional scripted browers datasets—often generated at negligible cost through scripted crawlers—LLM-based traffic simulation is considerably more expensive. Despite being far more scalable and efficient than recruiting human participants, generating high-quality interaction traces using commercial LLMs (e.g., via API-driven computer use agents) incurs both monetary and compute costs. As a result, even modest datasets with a few hundred samples per class can rapidly consume daily token budgets or GPU inference quotas.

This cost barrier presents a fundamental tradeoff: as websites become more dynamic and interaction-rich, effective fingerprinting requires data that captures complex user behavior. Scripted methods no longer suffice, and models increasingly demand training data that reflects semantic intent, exploratory actions, and user-specific navigation. This trend implies that future WFP systems must either rely on more powerful synthetic pipelines or be restricted to narrower threat models where such data is obtainable.

\paragraph{Scalability Challenges for Open-World Evaluation}

A practical consequence of this cost imbalance is the difficulty in scaling LLM-based data generation to open-world settings. Prior WFP benchmarks have explored datasets with thousands of websites—such as the 5k-site Tor set or the Undefendable corpus—but replicating such scale under an LLM-agent framework is currently prohibitive for most researchers without substantial GPU infrastructure or funding. As a result, evaluating WFP under realistic open-world conditions remains an unsolved challenge.

That said, trends in compute accessibility and inference pricing (e.g., through more efficient LLM architectures, quantization, and API competition) suggest this barrier may soon fall. In the near future, it may become feasible to generate diverse, high-volume synthetic traffic at web-scale, opening new possibilities for large-scale WFP benchmarking and defense testing.

\paragraph{Targeted Fingerprinting and the Rise of Personalized Simulation}

Interestingly, the precision and controllability of LLM agents open the door to a different attack vector: targeted fingerprinting. Unlike broad-scale open-world classification, attacks that aim to profile or deanonymize specific individuals may require only small amounts of data—provided the agent can model the target's browsing preferences, background, or goals. For example, if an attacker has partial knowledge of a user's demographics, profession, or daily routine, an LLM-driven simulation could potentially emulate their interaction pattern and improve classification accuracy against that individual.

Such personalized attacks were previously impractical due to data scarcity, but LLM conditioning capabilities make this increasingly feasible. We see this as a dual-use scenario: while it enhances attack realism, it also raises ethical and privacy concerns that must be addressed by the research community.

\section{Conclusion}
\label{sec:conclusion}

This work revisits Website Fingerprinting (WFP) under modern browsing conditions, where websites are dynamic, interaction-heavy, and personalized. We conduct a comprehensive evaluation across nine state-of-the-art WFP models using over 100 hours of real human browsing traces, revealing a consistent and significant performance gap between training on synthetic scripted traffic and evaluating on real users. Our results show that traditional non-interactive robot-generated datasets fail to capture the timing-sensitive and behavior-dependent traffic patterns induced by real-world usage.

Moreover, we find that user behavior introduces a surprising amount of intra-class variability—different individuals browsing the same site can produce traffic patterns that are distinct enough to confuse even strong classifiers. This highlights the importance of training datasets that are not only large, but also behaviorally diverse. In cross-user and cross-domain evaluations, even the most accurate models trained under idealized conditions suffer sharp performance degradation.

To address these challenges, we propose a scalable multi-agent data generation pipeline powered by large language models (LLMs). Our framework simulates user interactions by combining high-level decision agents with low-level execution agents based on LLM-driven browser control. Despite operating at just one-third the cost of real human data collection, this synthetic traffic significantly improves generalization: in many cases, models trained on LLM-generated traces outperform those trained on scripted data by over 3$\times$.

While promising, this approach still has limitations. Compared to near-free synthetic crawlers, multi-agent LLM generation is more costly and constrained by current API usage limits. Building massive open-world datasets with thousands of websites remains challenging without dedicated GPU infrastructure or open-source foundation models. Nonetheless, as token prices drop and model quality rises, scalable simulation will become increasingly accessible.

In the longer term, our framework also opens the door to new research directions. LLMs can mimic specific personas, enabling targeted fingerprinting attacks on individuals with limited training data. As LLMs evolve to incorporate multimodal understanding and long-term memory, they may unlock even richer browsing simulations—bridging the gap between controlled evaluation and real-world threat modeling.

In summary, this work demonstrates that the future of WFP will be shaped not only by advances in modeling, but by the ability to construct realistic, diverse, and scalable datasets. Our findings advocate for a paradigm shift in dataset design, one that embraces behavioral realism as a first-class requirement.

%%
%% The acknowledgments section is defined using the "acks" environment
%% (and NOT an unnumbered section). This ensures the proper
%% identification of the section in the article metadata, and the
%% consistent spelling of the heading.
% \begin{acks}
% To Robert, for the bagels and explaining CMYK and color spaces.
% \end{acks}

%%
%% The next two lines define the bibliography style to be used, and
%% the bibliography file.

% \bibliographystyle{ACM-Reference-Format}
% \bibliography{sample-base}

\bibliographystyle{IEEEtranS}
\bibliography{main}

% Generated by IEEEtranS.bst, version: 1.14 (2015/08/26)
\begin{thebibliography}{10}
\providecommand{\url}[1]{#1}
\csname url@samestyle\endcsname
\providecommand{\newblock}{\relax}
\providecommand{\bibinfo}[2]{#2}
\providecommand{\BIBentrySTDinterwordspacing}{\spaceskip=0pt\relax}
\providecommand{\BIBentryALTinterwordstretchfactor}{4}
\providecommand{\BIBentryALTinterwordspacing}{\spaceskip=\fontdimen2\font plus
\BIBentryALTinterwordstretchfactor\fontdimen3\font minus \fontdimen4\font\relax}
\providecommand{\BIBforeignlanguage}[2]{{%
\expandafter\ifx\csname l@#1\endcsname\relax
\typeout{** WARNING: IEEEtranS.bst: No hyphenation pattern has been}%
\typeout{** loaded for the language `#1'. Using the pattern for}%
\typeout{** the default language instead.}%
\else
\language=\csname l@#1\endcsname
\fi
#2}}
\providecommand{\BIBdecl}{\relax}
\BIBdecl

\bibitem{netclr}
\BIBentryALTinterwordspacing
A.~Bahramali, A.~Bozorgi, and A.~Houmansadr, ``Realistic website fingerprinting by augmenting network traces,'' in \emph{Proceedings of the 2023 ACM SIGSAC Conference on Computer and Communications Security}, ser. CCS '23.\hskip 1em plus 0.5em minus 0.4em\relax New York, NY, USA: Association for Computing Machinery, 2023, p. 1035–1049. [Online]. Available: \url{https://doi.org/10.1145/3576915.3616639}
\BIBentrySTDinterwordspacing

\bibitem{varcnn}
H.~Bhat, X.~Lu, M.~Lindorfer, N.~Feamster, and V.~Paxson, ``Var-cnn: A data-efficient website fingerprinting attack based on deep learning,'' in \emph{Proceedings of the Network and Distributed System Security Symposium (NDSS)}, 2019.

\bibitem{old_wfnet}
T.~Burns, C.~Song, I.~Seskar, J.~Ortiz, and R.~P. Martin, ``A simplified machine learning approach to classifying individual websites,'' in \emph{GLOBECOM 2022 - 2022 IEEE Global Communications Conference}, 2022, pp. 6109--6114.

\bibitem{cai2012touching}
X.~Cai, X.~Zhang, B.~Joshi, and R.~Johnson, ``Touching from a distance: Website fingerprinting attacks and defenses,'' in \emph{Proceedings of the 2012 ACM conference on Computer and communications security}.\hskip 1em plus 0.5em minus 0.4em\relax ACM, 2012, pp. 605--616.

\bibitem{CHEN2021108298}
\BIBentryALTinterwordspacing
M.~Chen, Y.~Wang, H.~Xu, and X.~Zhu, ``Few-shot website fingerprinting attack,'' \emph{Computer Networks}, vol. 198, p. 108298, 2021. [Online]. Available: \url{https://www.sciencedirect.com/science/article/pii/S1389128621003108}
\BIBentrySTDinterwordspacing

\bibitem{cheng2025holmes}
\BIBentryALTinterwordspacing
Y.~Cheng, Y.~Zhu, B.~Li, peishuai sun, Y.~Ding, X.~Deng, and Q.~Liu, ``{HOLMES} \& {WATSON}: A robust and lightweight {HTTPS} website fingerprinting through {HTTP} version parallelism,'' in \emph{THE WEB CONFERENCE 2025}, 2025. [Online]. Available: \url{https://openreview.net/forum?id=lSJ8VjjimZ}
\BIBentrySTDinterwordspacing

\bibitem{277132}
\BIBentryALTinterwordspacing
G.~Cherubin, R.~Jansen, and C.~Troncoso, ``Online website fingerprinting: Evaluating website fingerprinting attacks on tor in the real world,'' in \emph{31st USENIX Security Symposium (USENIX Security 22)}.\hskip 1em plus 0.5em minus 0.4em\relax Boston, MA: USENIX Association, Aug. 2022, pp. 753--770. [Online]. Available: \url{https://www.usenix.org/conference/usenixsecurity22/presentation/cherubin}
\BIBentrySTDinterwordspacing

\bibitem{online_website}
\BIBentryALTinterwordspacing
------, ``Online website fingerprinting: Evaluating website fingerprinting attacks on tor in the real world,'' in \emph{31st USENIX Security Symposium (USENIX Security 22)}.\hskip 1em plus 0.5em minus 0.4em\relax Boston, MA: USENIX Association, Aug. 2022, pp. 753--770. [Online]. Available: \url{https://www.usenix.org/conference/usenixsecurity22/presentation/cherubin}
\BIBentrySTDinterwordspacing

\bibitem{deng2024robustreliableearlystagewebsite}
\BIBentryALTinterwordspacing
X.~Deng, Q.~Li, and K.~Xu, ``Robust and reliable early-stage website fingerprinting attacks via spatial-temporal distribution analysis,'' 2024. [Online]. Available: \url{https://arxiv.org/abs/2407.00918}
\BIBentrySTDinterwordspacing

\bibitem{10179464}
X.~Deng, Q.~Yin, Z.~Liu, X.~Zhao, Q.~Li, M.~Xu, K.~Xu, and J.~Wu, ``Robust multi-tab website fingerprinting attacks in the wild,'' in \emph{2023 IEEE Symposium on Security and Privacy (SP)}, 2023, pp. 1005--1022.

\bibitem{ares}
\BIBentryALTinterwordspacing
X.~Deng, X.~Zhao, Q.~Yin, Z.~Liu, Q.~Li, M.~Xu, K.~Xu, and J.~Wu, ``Towards robust multi-tab website fingerprinting,'' 2025. [Online]. Available: \url{https://arxiv.org/abs/2501.12622}
\BIBentrySTDinterwordspacing

\bibitem{gong2020zero}
J.~Gong and T.~Wang, ``Zero-delay lightweight defenses against website fingerprinting,'' in \emph{USENIX Security Symposium}, 2020, pp. 717--734.

\bibitem{bapm}
\BIBentryALTinterwordspacing
Z.~Guan, G.~Xiong, G.~Gou, Z.~Li, M.~Cui, and C.~Liu, ``Bapm: Block attention profiling model for multi-tab website fingerprinting attacks on tor,'' in \emph{Proceedings of the 37th Annual Computer Security Applications Conference}, ser. ACSAC '21.\hskip 1em plus 0.5em minus 0.4em\relax New York, NY, USA: Association for Computing Machinery, 2021, p. 248–259. [Online]. Available: \url{https://doi.org/10.1145/3485832.3485891}
\BIBentrySTDinterwordspacing

\bibitem{ijcai2024p890}
\BIBentryALTinterwordspacing
T.~Guo, X.~Chen, Y.~Wang, R.~Chang, S.~Pei, N.~V. Chawla, O.~Wiest, and X.~Zhang, ``Large language model based multi-agents: A survey of progress and challenges,'' in \emph{Proceedings of the Thirty-Third International Joint Conference on Artificial Intelligence, {IJCAI-24}}, K.~Larson, Ed.\hskip 1em plus 0.5em minus 0.4em\relax International Joint Conferences on Artificial Intelligence Organization, 8 2024, pp. 8048--8057, survey Track. [Online]. Available: \url{https://doi.org/10.24963/ijcai.2024/890}
\BIBentrySTDinterwordspacing

\bibitem{hayes2016kfingerprinting}
J.~Hayes and G.~Danezis, ``k-fingerprinting: a robust scalable website fingerprinting technique,'' in \emph{USENIX Security}, 2016.

\bibitem{hu2025positionresponsiblellmempoweredmultiagent}
\BIBentryALTinterwordspacing
J.~Hu, Y.~Dong, S.~Ao, Z.~Li, B.~Wang, L.~Singh, G.~Cheng, S.~D. Ramchurn, and X.~Huang, ``Position: Towards a responsible llm-empowered multi-agent systems,'' 2025. [Online]. Available: \url{https://arxiv.org/abs/2502.01714}
\BIBentrySTDinterwordspacing

\bibitem{explainwf-popets2023}
R.~Jansen and R.~Wails, ``Data-explainable website fingerprinting with network simulation,'' \emph{Proceedings on Privacy Enhancing Technologies}, vol. 2023, no.~4, 2023, see also \url{https://explainwf-popets2023.github.io}.

\bibitem{tmwf}
\BIBentryALTinterwordspacing
Z.~Jin, T.~Lu, S.~Luo, and J.~Shang, ``Transformer-based model for multi-tab website fingerprinting attack,'' in \emph{Proceedings of the 2023 ACM SIGSAC Conference on Computer and Communications Security}, ser. CCS '23.\hskip 1em plus 0.5em minus 0.4em\relax New York, NY, USA: Association for Computing Machinery, 2023, p. 1050–1064. [Online]. Available: \url{https://doi.org/10.1145/3576915.3623107}
\BIBentrySTDinterwordspacing

\bibitem{critical_evaluation}
\BIBentryALTinterwordspacing
M.~Juarez, S.~Afroz, G.~Acar, C.~Diaz, and R.~Greenstadt, ``A critical evaluation of website fingerprinting attacks,'' in \emph{Proceedings of the 2014 ACM SIGSAC Conference on Computer and Communications Security}, ser. CCS '14.\hskip 1em plus 0.5em minus 0.4em\relax New York, NY, USA: Association for Computing Machinery, 2014, p. 263–274. [Online]. Available: \url{https://doi.org/10.1145/2660267.2660368}
\BIBentrySTDinterwordspacing

\bibitem{juarez2016towards}
------, ``Towards an efficient website fingerprinting defense,'' in \emph{European Symposium on Research in Computer Security}.\hskip 1em plus 0.5em minus 0.4em\relax Springer, 2016, pp. 27--46.

\bibitem{juarez2016critical}
M.~Juarez and et~al., ``A critical evaluation of website fingerprinting attacks,'' in \emph{Proceedings of the 2016 ACM SIGSAC Conference on Computer and Communications Security}, 2016, pp. 263--274.

\bibitem{Li_Wang_Zeng_Wu_Yang_2024}
\BIBentryALTinterwordspacing
X.~Li, S.~Wang, S.~Zeng, Y.~Wu, and Y.~Yang, ``\BIBforeignlanguage{en}{A survey on llm-based multi-agent systems: workflow, infrastructure, and challenges},'' Oct. 2024. [Online]. Available: \url{http://dx.doi.org/10.1007/s44336-024-00009-2}
\BIBentrySTDinterwordspacing

\bibitem{ref:webarena}
H.~e.~a. Liu, ``Webarena: A realistic web environment for building autonomous agents,'' in \emph{NeurIPS}, 2023.

\bibitem{agentbench}
\BIBentryALTinterwordspacing
X.~Liu, H.~Yu, H.~Zhang, Y.~Xu, X.~Lei, H.~Lai, Y.~Gu, H.~Ding, K.~Men, K.~Yang, S.~Zhang, X.~Deng, A.~Zeng, Z.~Du, C.~Zhang, S.~Shen, T.~Zhang, Y.~Su, H.~Sun, M.~Huang, Y.~Dong, and J.~Tang, ``Agentbench: Evaluating llms as agents,'' 2023. [Online]. Available: \url{https://arxiv.org/abs/2308.03688}
\BIBentrySTDinterwordspacing

\bibitem{298090}
\BIBentryALTinterwordspacing
A.~Mitseva and A.~Panchenko, ``Stop, don{\textquoteright}t click here anymore: Boosting website fingerprinting by considering sets of subpages,'' in \emph{33rd USENIX Security Symposium (USENIX Security 24)}.\hskip 1em plus 0.5em minus 0.4em\relax Philadelphia, PA: USENIX Association, Aug. 2024, pp. 4139--4156. [Online]. Available: \url{https://www.usenix.org/conference/usenixsecurity24/presentation/mitseva}
\BIBentrySTDinterwordspacing

\bibitem{oh2021online}
S.~E. Oh and A.~Houmansadr, ``Online website fingerprinting: Evaluating website fingerprinting attacks on real-world browsing behavior,'' in \emph{Proceedings of the 2021 ACM SIGSAC Conference on Computer and Communications Security (CCS)}.\hskip 1em plus 0.5em minus 0.4em\relax ACM, 2021, pp. 2826--2841.

\bibitem{ohnishi2023subverting}
N.~Ohnishi and S.~J. Murdoch, ``Subverting website fingerprinting defenses with robust traffic representation,'' in \emph{NDSS}, 2023.

\bibitem{panchenko2011website}
A.~Panchenko, F.~Lanze, A.~Zinnen \emph{et~al.}, ``Website fingerprinting in onion routing based anonymization networks,'' in \emph{ACM WPES}, 2011.

\bibitem{tiktok}
M.~Rahman and M.~Wright, ``Tik-tok: The utility of packet timing in website fingerprinting attacks,'' \emph{Proceedings on Privacy Enhancing Technologies}, vol. 2020, no.~3, pp. 5--24, 2020.

\bibitem{sager2025aiagentscomputeruse}
\BIBentryALTinterwordspacing
P.~J. Sager, B.~Meyer, P.~Yan, R.~von Wartburg-Kottler, L.~Etaiwi, A.~Enayati, G.~Nobel, A.~Abdulkadir, B.~F. Grewe, and T.~Stadelmann, ``Ai agents for computer use: A review of instruction-based computer control, gui automation, and operator assistants,'' 2025. [Online]. Available: \url{https://arxiv.org/abs/2501.16150}
\BIBentrySTDinterwordspacing

\bibitem{tf}
\BIBentryALTinterwordspacing
P.~Sirinam, N.~Mathews, M.~S. Rahman, and M.~Wright, ``Triplet fingerprinting: More practical and portable website fingerprinting with n-shot learning,'' in \emph{Proceedings of the 2019 ACM SIGSAC Conference on Computer and Communications Security}, ser. CCS '19.\hskip 1em plus 0.5em minus 0.4em\relax New York, NY, USA: Association for Computing Machinery, 2019, p. 1131–1148. [Online]. Available: \url{https://doi.org/10.1145/3319535.3354217}
\BIBentrySTDinterwordspacing

\bibitem{sirinam2018systematic}
P.~Sirinam, M.~Juarez, J.~Hayes, and C.~Diaz, ``A systematic approach to developing and evaluating website fingerprinting defenses,'' in \emph{NDSS}, 2018.

\bibitem{df}
P.~Sirinam and et~al., ``Deep fingerprinting: Undermining website fingerprinting defenses with deep learning,'' in \emph{Proceedings of the 2018 ACM CCS}, 2018.

\bibitem{wfnet}
\BIBentryALTinterwordspacing
C.~Song, Z.~Fan, H.~Wang, and R.~Martin, ``Seamless website fingerprinting in multiple environments,'' 2024. [Online]. Available: \url{https://arxiv.org/abs/2407.19365}
\BIBentrySTDinterwordspacing

\bibitem{raptor2017early}
Y.~Sun, A.~Edmundson, L.~Vanbever, and N.~Feamster, ``Raptor: Routing attacks on privacy in tor,'' in \emph{USENIX Security Symposium}, 2015, pp. 271--286.

\bibitem{talebirad2023multiagentcollaborationharnessingpower}
\BIBentryALTinterwordspacing
Y.~Talebirad and A.~Nadiri, ``Multi-agent collaboration: Harnessing the power of intelligent llm agents,'' 2023. [Online]. Available: \url{https://arxiv.org/abs/2306.03314}
\BIBentrySTDinterwordspacing

\bibitem{trafficSliver2021}
G.~Wang and T.~Wang, ``Trafficsliver: Fighting website fingerprinting attacks with traffic splitting,'' \emph{Proceedings on Privacy Enhancing Technologies}, 2021.

\bibitem{llm_survey}
\BIBentryALTinterwordspacing
L.~Wang, C.~Ma, X.~Feng, Z.~Zhang, H.~Yang, J.~Zhang, Z.~Chen, J.~Tang, X.~Chen, Y.~Lin, W.~X. Zhao, Z.~Wei, and J.~Wen, ``A survey on large language model based autonomous agents,'' \emph{Frontiers of Computer Science}, vol.~18, no.~6, Mar. 2024. [Online]. Available: \url{http://dx.doi.org/10.1007/s11704-024-40231-1}
\BIBentrySTDinterwordspacing

\bibitem{tor_wang}
\BIBentryALTinterwordspacing
T.~Wang and I.~Goldberg, ``Improved website fingerprinting on tor,'' in \emph{Proceedings of the 12th ACM Workshop on Workshop on Privacy in the Electronic Society}, ser. WPES '13.\hskip 1em plus 0.5em minus 0.4em\relax New York, NY, USA: Association for Computing Machinery, 2013, p. 201–212. [Online]. Available: \url{https://doi.org/10.1145/2517840.2517851}
\BIBentrySTDinterwordspacing

\bibitem{wang2017walkie}
------, ``Walkie-talkie: An efficient defense against passive website fingerprinting attacks,'' in \emph{27th {USENIX} Security Symposium ({USENIX} Security 18)}, 2018.

\bibitem{wshop}
\BIBentryALTinterwordspacing
S.~Yao, H.~Chen, J.~Yang, and K.~Narasimhan, ``Webshop: Towards scalable real-world web interaction with grounded language agents,'' 2023. [Online]. Available: \url{https://arxiv.org/abs/2207.01206}
\BIBentrySTDinterwordspacing

\bibitem{10.1145/3658644.3690211}
\BIBentryALTinterwordspacing
X.~Zhao, X.~Deng, Q.~Li, Y.~Liu, Z.~Liu, K.~Sun, and K.~Xu, ``Towards fine-grained webpage fingerprinting at scale,'' in \emph{Proceedings of the 2024 on ACM SIGSAC Conference on Computer and Communications Security}, ser. CCS '24.\hskip 1em plus 0.5em minus 0.4em\relax New York, NY, USA: Association for Computing Machinery, 2024, p. 423–436. [Online]. Available: \url{https://doi.org/10.1145/3658644.3690211}
\BIBentrySTDinterwordspacing

\end{thebibliography}

%%
%% If your work has an appendix, this is the place to put it.
% \appendix

\end{document}